\documentclass{nature}

\usepackage[utf8]{inputenc}
\usepackage{graphicx}
\usepackage{aa_sym}
\usepackage{color}

\newcommand{\subs}[1]{\mbox{\scriptsize #1}}
\newcommand{\dndz}{\mbox{$\mathrm{d}n/\mathrm{d}z$}}
\newcommand{\dndx}{\mbox{$\mathrm{d}n/\mathrm{d}X$}}

\title{Nearly 100\% of the sky is covered by Lyman-$\mathbf{\alpha}$ emission around high redshift galaxies \protect\\
{\normalsize to be published in Nature}\protect\\
{\normalsize -- Preprint, authors' version}
}

\author{L. Wisotzki$^{1}$, R. Bacon$^{2}$, J. Brinchmann$^{3,4}$, S. Cantalupo$^{5}$, P. Richter$^{6}$, J. Schaye$^{3}$, K. B. Schmidt$^{1}$, T. Urrutia$^{1}$, P. M. Weilbacher$^{1}$, M. Akhlaghi$^{2}$, N. Bouch\'{e}$^{7}$, T. Contini$^{7}$, B. Guiderdoni$^{2}$, E. C. Herenz$^{8}$, H. Inami$^{2}$, J. Kerutt$^{1}$, F. Leclercq$^{2}$, R. A. Marino$^{5}$, M. Maseda$^{3}$, A. Monreal-Ibero$^{9, 10}$, T. Nanayakkara$^{3}$, J. Richard$^{2}$, R. Saust$^{1}$, M. Steinmetz$^{1}$, M. Wendt$^{1,6}$}

\begin{document}
\setlength{\unitlength}{1mm}
\sloppy

\maketitle

{\footnotesize
\begin{affiliations}
 \item Leibniz-Institut f\"ur Astrophysik Potsdam (AIP), An der Sternwarte 16, 14482 Potsdam, Germany
 \item Univ Lyon, Univ Lyon1, Ens de Lyon, CNRS, Centre de Recherche Astrophysique de Lyon UMR5574, F-69230, Saint-Genis-Laval, France
 \item Leiden Observatory, Leiden University, PO Box 9513, 2300 RA Leiden, The Netherlands
 \item Instituto de Astrof\'{i}sica e Ci\^{e}ncias do Espa\c{c}o, Universidade do Porto, CAUP, rua das Estrelas, 4150-762 Porto, Portugal
 \item Department of Physics, ETH Z\"{u}rich, CH-8093 Z\"{u}rich, Switzerland
 \item Institut f\"{u}r Physik und Astronomie, Universit\"{a}t Potsdam, 14476 Potsdam, Germany
 \item Institut de Recherche en Astrophysique et Plan\'{e}tologie (IRAP), Universit\'{e} de Toulouse, CNRS, UPS, 31400 Toulouse, France
 \item Department of Astronomy, Stockholm University, AlbaNova University Centre, 10691 Stockholm, Sweden
 \item Instituto de Astrof\'{i}sica de Canarias (IAC), 38205 La Laguna, Tenerife, Spain
 \item Universidad de La Laguna, Dpto.\ Astrof\'{i}sica, 38206 La Laguna, Tenerife, Spain
\end{affiliations}
}

\begin{abstract}
Galaxies are surrounded by large reservoirs of gas, mostly hydrogen, fed by inflows from the intergalactic medium and by outflows due to galactic winds. Absorption-line measurements along the sightlines to bright and rare background quasars indicate that this circumgalactic medium pervades far beyond the extent of starlight in galaxies, but very little is known about the spatial distribution of this gas. A new window into circumgalactic environments was recently opened with the discovery of ubiquitous extended Lyman-$\alpha$ emission from hydrogen around high-redshift galaxies\cite{Wisotzki:2016hw,Leclercq:2017cp}, facilitated by the extraordinary sensitivity of the MUSE instrument at the ESO Very Large Telescope\cite{Bacon:2015eh}. Due to the faintness of this emission, such measurements were previously limited to especially favourable systems\cite{Hayes:2011bd,Cantalupo:2014ig,Hennawi:2015ji} or to massive statistical averaging\cite{Steidel:2011jk,Momose:2014fe}. Here we demonstrate that low surface brightness Lyman-$\alpha$ emission surrounding faint galaxies at redshifts between 3 and 6 adds up to a projected sky coverage of nearly 100\%. The corresponding rate of incidence (the mean number of Lyman-$\alpha$ emitters penetrated by any arbitrary line of sight) is well above unity and similar to the incidence rate of high column density absorbers frequently detected in the spectra of distant quasars\cite{Peroux:2003eg,Songaila:2010ft,Zafar:2013ew,Crighton:2015jc}. This similarity suggests that most circumgalactic atomic hydrogen at these redshifts has now been detected also in emission. 
\end{abstract}

\clearpage

The Lyman-$\alpha$ transition of atomic hydrogen at 121.6\,nm (short: Ly$\alpha$) is an important tracer of warm ($\sim 10^4$~K) gas in and around galaxies, especially at cosmological redshifts $z>2$ where the line becomes observable with ground-based observatories. Tracing cosmic hydrogen through its Ly$\alpha$ emission has been a long-standing goal of observational astrophysics\cite{Partridge:1967iw,Hogan:1987wh,Gould:1996gi}, but the extremely low surface brightness of spatially extended Ly$\alpha$ emission is a formidable obstacle in this quest. An important technological step forward in this respect has been afforded by the Multi-Unit Spectroscopic Explorer (MUSE), developed by our team and installed in 2014 at the ESO Very Large Telescope\cite{Bacon:2010jn,Bacon:2014wp}. MUSE was specifically designed for maximum sensitivity to Ly$\alpha$ emission in the redshift range 2.9–6.6, a key era in the formation and evolution of galaxies. We performed very long MUSE exposures in two fields that were previously mapped to extreme depths with the Hubble Space Telescope (HST): the Hubble Deep Field South\cite{Casertano:2000fn} and the Hubble Ultra Deep Field\cite{Beckwith:2006hp}. In our MUSE data we detected 270 Ly$\alpha$ emitting galaxies at $3<z<6$ (Methods), many of which are barely visible with HST. Extended Ly$\alpha$ haloes around these galaxies can be traced to distances of several arcseconds from the source centres\cite{Wisotzki:2016hw,Leclercq:2017cp}. 

In a first approach to estimate the integrated cross-section of extended Ly$\alpha$ emission we constructed redshift-integrated Ly$\alpha$ maps in the two observed fields as follows (Methods): We extracted pseudo-narrowband Ly$\alpha$ subimages around each object from the MUSE data. Co-adding all subimages over the full redshift range followed by some spatial filtering yielded a Ly$\alpha$ image for each field. Fig.~1 shows this image for the HUDF. Counting the number of pixels above a given Ly$\alpha$ surface brightness $s_{\subs{Ly$\alpha$}}$ yields the fractional sky coverage $f_{\subs{Ly$\alpha$}}$. For a threshold of $s_{\subs{Ly$\alpha$}} > 10^{-19}$ erg s$^{-1}$ cm$^{-2}$ arcsec$^{-2}$ -- the typical limiting surface brightness in the narrowband images -- we find a sky coverage of 46\% in the HUDF and 45\% in the HDFS (Methods). While this result already suggests that the sky coverage might further increase for even lower thresholds, the approach is hampered by noise and the need to apply spatial filtering.

To lower the surface brightness limit beyond the sensitivity of individual sightlines, we employed a combination of azimuthal averaging and image stacking (Methods). We first computed radial Ly$\alpha$ surface brightness profiles for each object by averaging over concentric annuli. Motivated by the fact that our Ly$\alpha$ halo profiles do not, on average, depend strongly on Ly$\alpha$ luminosity\cite{Leclercq:2017cp}, we then median-combined the individual images in three redshift bins (3--4, 4--5, 5--6), and approximated the radial profiles of the median images by smooth fitting functions. Fig.~2 illustrates this process. Rescaled to the actual Ly$\alpha$ fluxes of our objects, the median-stacked profiles reach surface brightness levels of $s_{\subs{Ly$\alpha$, lim}} \simeq (5, 4, 4) \times 10^{-21}$ erg s$^{-1}$ cm$^{-2}$ arcsec$^{-2}$ ($1\sigma$), an order of magnitude below the limit achievable for single sightlines. 

From the scaled median-stacked profiles we created synthetic Ly$\alpha$ maps for the three redshift bins and for the full redshift range. These maps, shown in Fig.~3a, represent idealised, noise-free and seeing-corrected models of the Ly$\alpha$ distribution in the sky. The resulting cumulative fractional Ly$\alpha$ sky coverage $f_{\subs{Ly$\alpha$}}$ is shown in Fig.~3b as a function of the surface brightness threshold. At $s_{\subs{Ly$\alpha$}} \simeq 10^{-20}$ erg s$^{-1}$ cm$^{-2}$ arcsec$^{-2}$, $f_{\subs{Ly$\alpha$}}$ is already well above 80\% and still increasing. At the faintest levels probed by our data, the values of $f_{\subs{Ly$\alpha$}}$ formally add up to more than 100\%, a clear sign that Ly$\alpha$ emission regions at different redshifts substantially overlap in projection.

The sky coverage is an intuitively appealing number but of limited use as it saturates at 100\%. A closely related but more physically useful quantity is the incidence rate \dndz, the average number of Ly$\alpha$ emitting regions per unit redshift passed by a typical line of sight, for a given surface brightness level. This quantity can be directly compared to \dndz\ obtained from absorption line statistics, for different absorber column densities. We also corrected for cosmological surface brightness dimming by moving from observed surface brightness $s_{\subs{Ly$\alpha$}}$ to intrinsic `surface luminosity' $S_{\subs{Ly$\alpha$}}$, expressed in erg s$^{-1}$ kpc$^{-2}$. Furthermore, we accounted for the inevitable faint-end incompleteness of the Ly$\alpha$ emitter sample by tying the integration to a completeness-corrected population distribution statistic (Methods). The resulting cumulative incidence rates as functions of surface luminosity threshold are presented in Fig.~4a. Values of $\dndz >1$ indicate that a random sightline passes on average through more than one emitter within a redshift interval of $\Delta z = 1$.
In Fig.~4b we compare our measured incidence rates with the statistics of atomic hydrogen detected in absorption against background quasars\cite{Peroux:2003eg,Songaila:2010ft,Zafar:2013ew,Crighton:2015jc}. We find that emission and absorption incidence rates \dndz$_\mathrm{em}$ and \dndz$_\mathrm{abs}$ have a similar range of values, which we use to tentatively match surface luminosities to column densities.

Emission regions with $\log_{10} S_{\subs{Ly$\alpha$}}/\mathrm{erg}\: \mathrm{s}^{-1}\:\mathrm{kpc}^{-2} \ga 38$ (for brevity we omit the units in the following), typically at radial distances $\la 2''$, have a \dndz$_\mathrm{em}$ of $\sim$0.5 per unit redshift which is comparable to Damped Ly$\alpha$ Absorbers\cite{Peroux:2003eg,Crighton:2015jc} (DLAs) with column densities of $\log_{10} N(\mbox{\ion{H}{i}})/\mathrm{cm}^{-2} > 20.3$. At redshifts $z\la 3.5$ this result broadly agrees with the findings in ref.~\cite{Rauch:2008jy} (based on long-slit spectroscopy of a much smaller sample), marked by the blue crosses in Fig.~4. Our data also show that the trend of \dndz\ with redshift is very similar for absorbers and emitters. It is thus suggestive to identify DLAs with Ly$\alpha$-emitting regions at levels of $\log_{10} S_{\subs{Ly$\alpha$}} > 38$, which is also approximately the limit for the detection of individual Ly$\alpha$ haloes\cite{Wisotzki:2016hw,Leclercq:2017cp}. These photons likely originate in star-forming regions and are then resonantly scattered outwards\cite{Laursen:2007kl,Barnes:2010ew,Verhamme:2012kb}, possibly enhanced by cooling radiation during the accretion of gas into dark matter haloes\cite{Haiman:2000di,Fardal:2001ds,Furlanetto:2005jq}.

Moving to lower column densities of atomic hydrogen, Fig.~4 shows that systems with $\log_{10} N(\mbox{\ion{H}{i}})/\mathrm{cm}^{-2} > 19$ (sometimes called Sub-DLAs or SDLAs) and Ly$\alpha$ emission regions with $\log_{10} S_{\subs{Ly$\alpha$}} \ga 37.5$ both have \dndz\ of order unity. At even lower $N(\mbox{\ion{H}{i}})$, Lyman Limit Systems (LLSs) with $\log_{10} N(\mbox{\ion{H}{i}})/\mathrm{cm}^{-2} > 17$ give rise to several incidences per unit redshift\cite{Songaila:2010ft}, which can be approximately matched in emission by surface luminosities of $\log_{10} S_{\subs{Ly$\alpha$}} \ga 37$, a level just detectable in our median stacks. Above column densities of $\sim 10^{18}$~cm$^{-2}$ hydrogen becomes self-shielded to ionising radiation and thus at least partly atomic\cite{Altay:2011iu}. The approximate equality of the incidence rates of high column density \ion{H}{i} absorbers and very low surface brightness Ly$\alpha$-emitting regions therefore suggests that in these regions we observe the faint glow of circumgalactic atomic hydrogen. Taking into account systematic errors introduces an uncertainty of a factor $\sim$2 (Methods), but these uncertainties do not affect our  conclusion that most atomic hydrogen at these redshifts is now detected also in emission.

Our results suggest that \dndz$_\mathrm{em}$ increases mildly with redshift. In Fig.~4b we compare our measurements with the expected redshift dependence for an intrinsically non-evolving population of emitters. Such a population can be described by a constant incidence rate \dndx$_\mathrm{em}$ per comoving path length $X$ along the line of sight, a quantity commonly used in quasar absorption line studies. Within the redshift range considered here, a constant \dndx$_\mathrm{em}(z)$ translates into a \dndz$_\mathrm{em}(z)$ similar to the trend suggested by our data points. Since the Ly$\alpha$ luminosity function also shows very little evolution within the redshift range of our sample\cite{Drake:2017jq}, we conclude that the circumgalactic Ly$\alpha$ emission properties do not change much between redshifts 6 and 3. Nevertheless, given the error bars the data are also compatible with a constant \dndz$_\mathrm{em}(z)$, corresponding to a modest decrease of incidence rates per comoving path length.

What powers this low-level Ly$\alpha$ emission that we tentatively identify as originating from circumgalactic Lyman limit systems? At large galactocentric distances, Ly$\alpha$ fluorescence of optically thick gas excited by the cosmic UV background (UVB) becomes a viable possibility\cite{Gould:1996gi}. The expected fluorescent Ly$\alpha$ surface brightness of a Lyman Limit System at $z=3$ was calculated in ref.~\cite{Cantalupo:2005hq} and recently updated for $z=3.5$ in ref.~\cite{Gallego:2018dj} as $s_{\subs{Ly$\alpha$,UVB}} \simeq 1.1\times 10^{-20}$ erg s$^{-1}$ cm$^{-2}$ arcsec$^{-2}$, just about within the sensitivity range of our stacked data and consistent with the marginal signal at large radii in our lowest redshift bin. This suggests that at least some of the extremely faint Ly$\alpha$ emission detected by MUSE may be due to this omnipresent glow, opening a new window into a significant but previously invisible component of the cosmic matter distribution.

\begin{addendum}
 \item All authors wish to thank ESO staff for their support that made these observations possible. 
 L.W., J.K., R.S. and T.U. acknowledge support by the Competitive Fund of the Leibniz Association through grants SAW-2013-AIP-4 and SAW-2015-AIP-2.
 R.B., H.I., F.L. and M.A. are supported by the ERC advanced grant 339659-MUSICOS. 
J.B. acknowledges support by FCT grants UID/FIS/04434/2013 and IF/01654/2014/CP1215/CT0003 and by FEDER through COMPETE2020 (POCI-01-0145-FEDER-007672).
 J.R. acknowledges support from the ERC starting grant 336736-CALENDS.
 P.M.W. received support through BMBF Verbundforschung, grant 05A17BAA.
 T.C., N.B. and B.G. acknowledge support by ANR FOGHAR (ANR-13-BS05-0010-02).
 T.C. and N.B. were also supported by OCEVU Labex (ANR-11-LABX-0060), and by the A*MIDEX project (ANR-11-IDEX-0001-02) funded by the “Investissements d’avenir” French government program. 
 A.M.I. acknowledges support from MINECO through project AYA2015-68217-P.
S.C. acknowledges support from Swiss National Science Foundation grant PP00P2\_163824.
 
 \item[Author contributions] L.W. conceived the project. 
R.B. led the data acquisition and data reduction. 
R.B., J.B., E.C.H, H.I., J.K., K.B.S., T.U. and L.W. developed and performed the sample selection. 
L.W. analysed the data, with input by R.B., J.B. and P.M.W. 
S.C., P.R., J.S., M.S. and L.W. worked on the interpretation of the results. 
L.W. wrote the manuscript and produced the figures, with K.B.S. contributing to their design.
All coauthors provided critical feedback to the text and helped shape the manuscript.

\item[Author information] Reprints and permissions information is available at www.nature.com/reprints. The authors declare no competing financial interests. Readers are welcome to comment on the online version of the paper. Publisher’s note: Springer Nature remains neutral with regard to jurisdictional claims in published maps and institutional affiliations. Correspondence and requests for materials should be addressed to L.W. (lwisotzki@aip.de).


\end{addendum}

\clearpage

\section*{References} 

\clearpage

\renewcommand{\figurename}{\textbf{Fig.}}

\begin{figure}
\includegraphics[width=\textwidth]{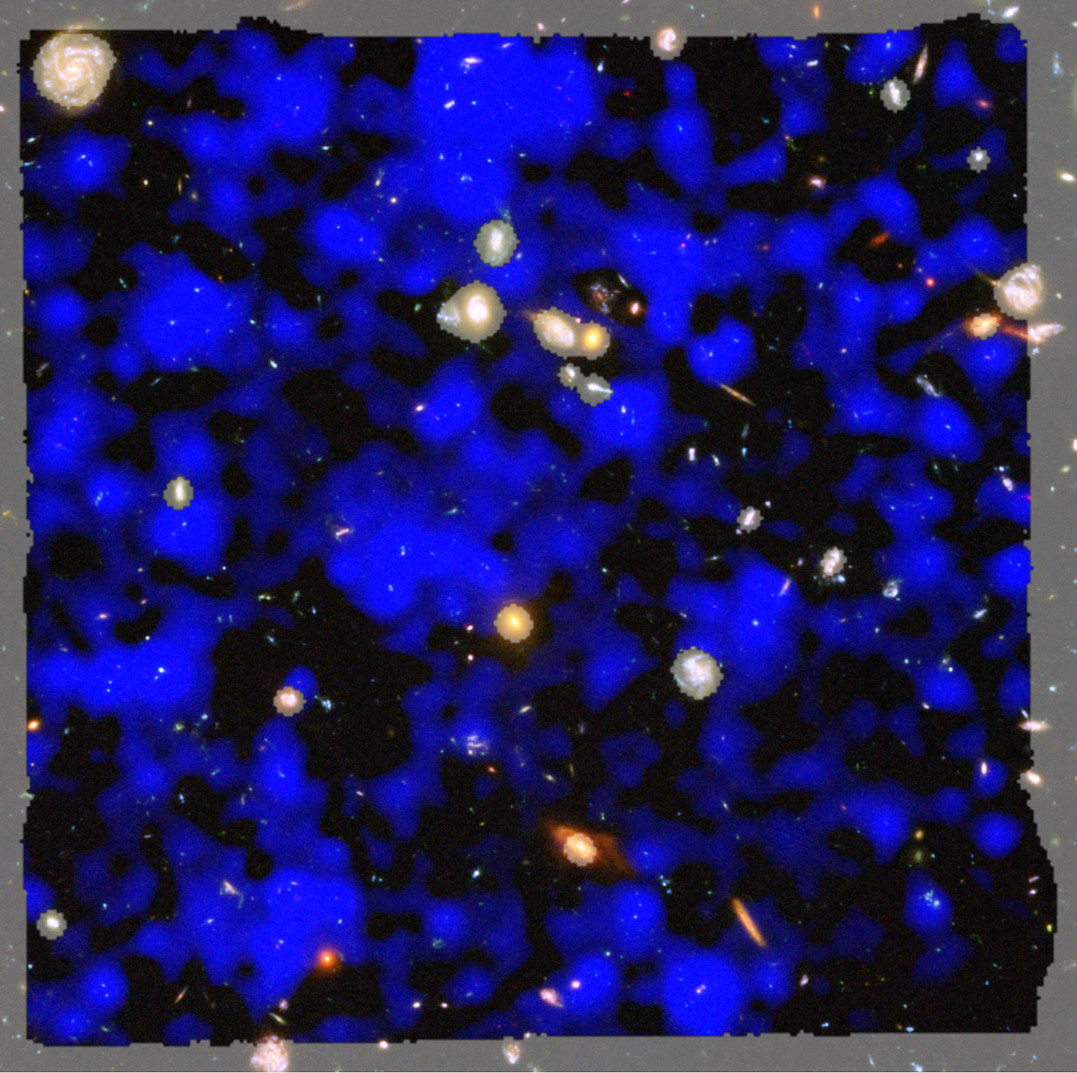}
\caption[]{\textbf{Distribution of the observed Lyman-$\alpha$ emission in the Hubble Ultra Deep Field.} The underlying image is a colour composite obtained by the Hubble Space Telescope\cite{Beckwith:2006hp}, restricted to the $1'\times 1'$ section observed with MUSE. The extended Ly$\alpha$ emission detected by MUSE is superimposed in blue, summed over the redshift range $3 < z < 6$ and spatially filtered to suppress the noise. The grey semi-transparent areas outline the MUSE field of view and mask the brightest foreground galaxies.  The dynamic range of the Ly$\alpha$ overlay was adjusted such that the faintest visible structures have a surface brightness of $10^{-19}$ erg s$^{-1}$ cm$^{-2}$ arcsec$^{-2}$. 
}
\end{figure}

\clearpage

\begin{figure}
\includegraphics[width=\textwidth]{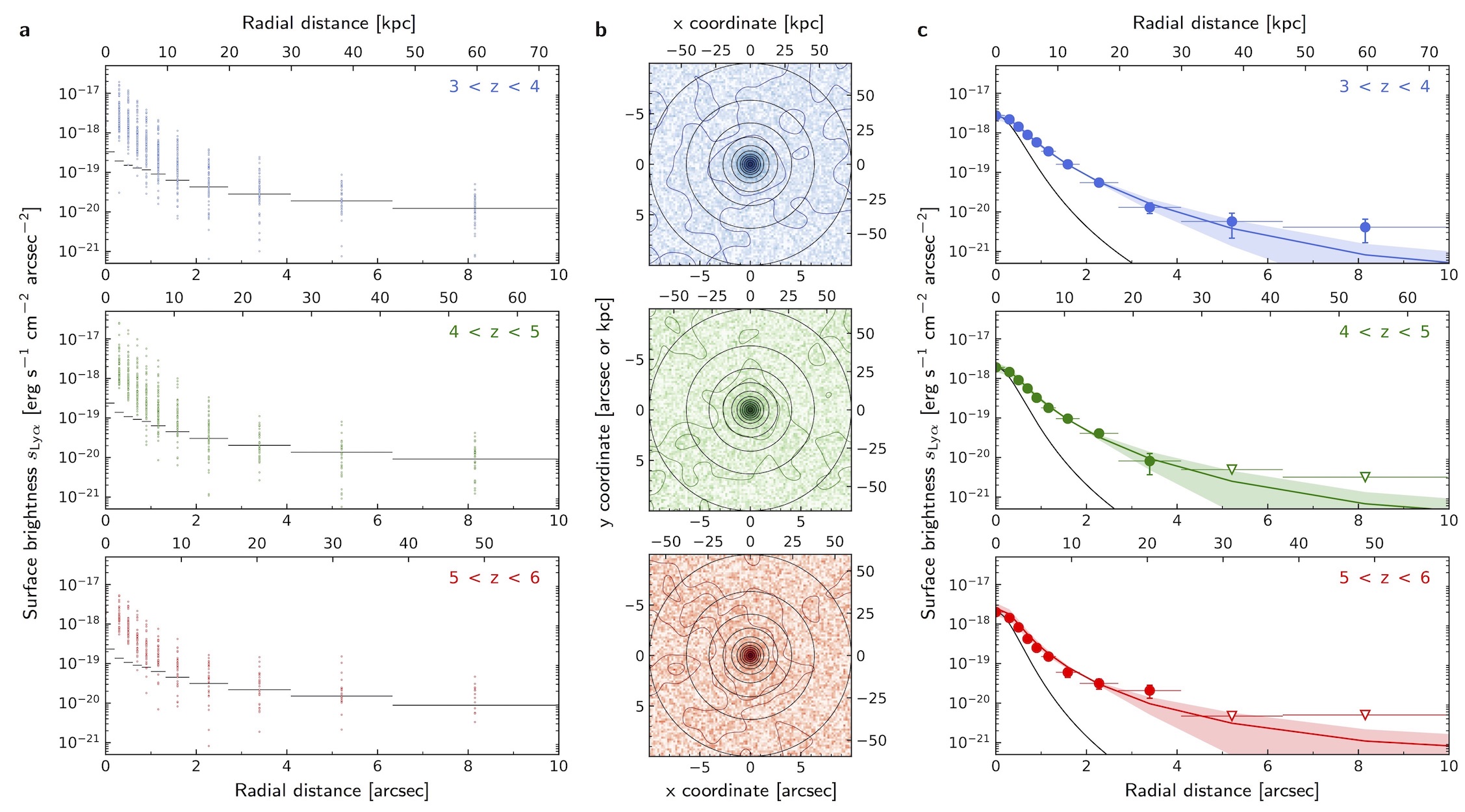}
\caption[]{\textbf{Stacking and construction of representative radial Lyman-$\alpha$ profiles.}  \textbf{a,} Ly$\alpha$ surface brightness profiles of the individual galaxies in the HUDF and HDFS, azimuthally averaged in concentric annuli. Horizontal bars specify the widths of the annuli. The three panels represent three disjoint redshift ranges as indicated. Radial coordinates are given in angular (bottom axes) as well as physical (top axes) units, the latter evaluated at the centre of each redshift range assuming a cosmological model with $h = 0.7$, $\Omega_{\subs{m}} = 0.3$ and $\Omega_\Lambda = 0.7$. \textbf{b,} Median-stacked Ly$\alpha$ images for these redshift bins. The contours trace surface brightnesses of $(0.5, 2)\times 10^{-20}$ erg s$^{\mathsf{-1}}$ cm$^{\mathsf{-2}}$ arcsec$^{\mathsf{-2}}$ after subtracting a model image, smoothing the residual with a Gaussian of 2\arcsec\ FWHM and adding back the model. The overplotted circles show the boundaries of the annuli used to extract the radial profiles. \textbf{c,} Azimuthally averaged radial profiles of the median-stacked images. The vertical bars on the data points quantify the 1$\sigma$ errors within each annulus, while the horizontal bars again indicate the widths of the annuli. The black line in each panel traces the radial shape of a scaled point source, demonstrating that the median Ly$\alpha$ emission is well-resolved for radii $\ga 1$ arcsec. The solid, coloured curves show the extracted profiles from 2-dimensional surface brightness model fits to the images, with the shaded regions indicating the estimated 1$\sigma$ uncertainties of the fits. }
\end{figure}

\clearpage

\begin{figure}
\includegraphics[width=0.5\textwidth]{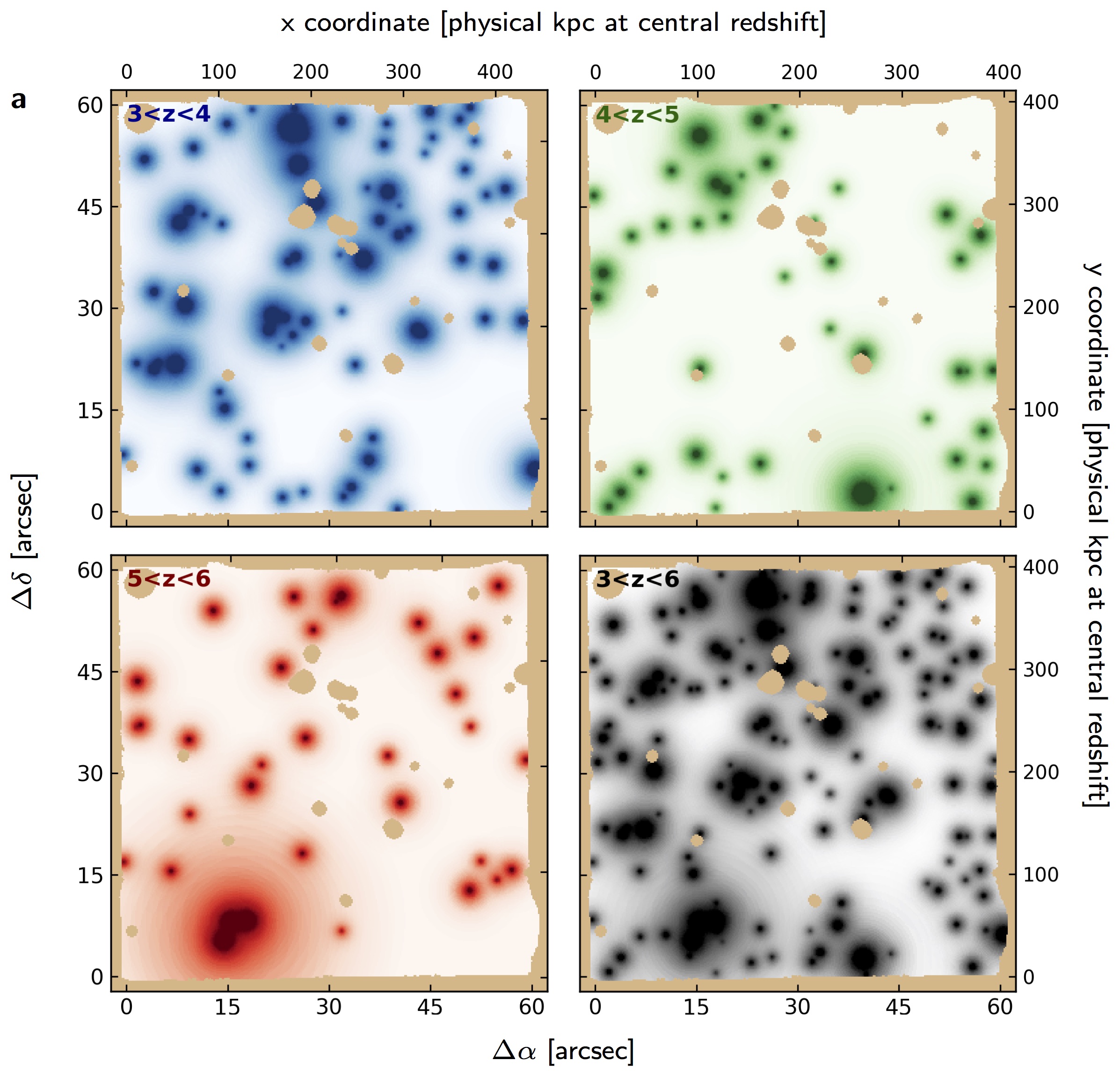}
\hspace{\fill}
\includegraphics[width=0.5\textwidth]{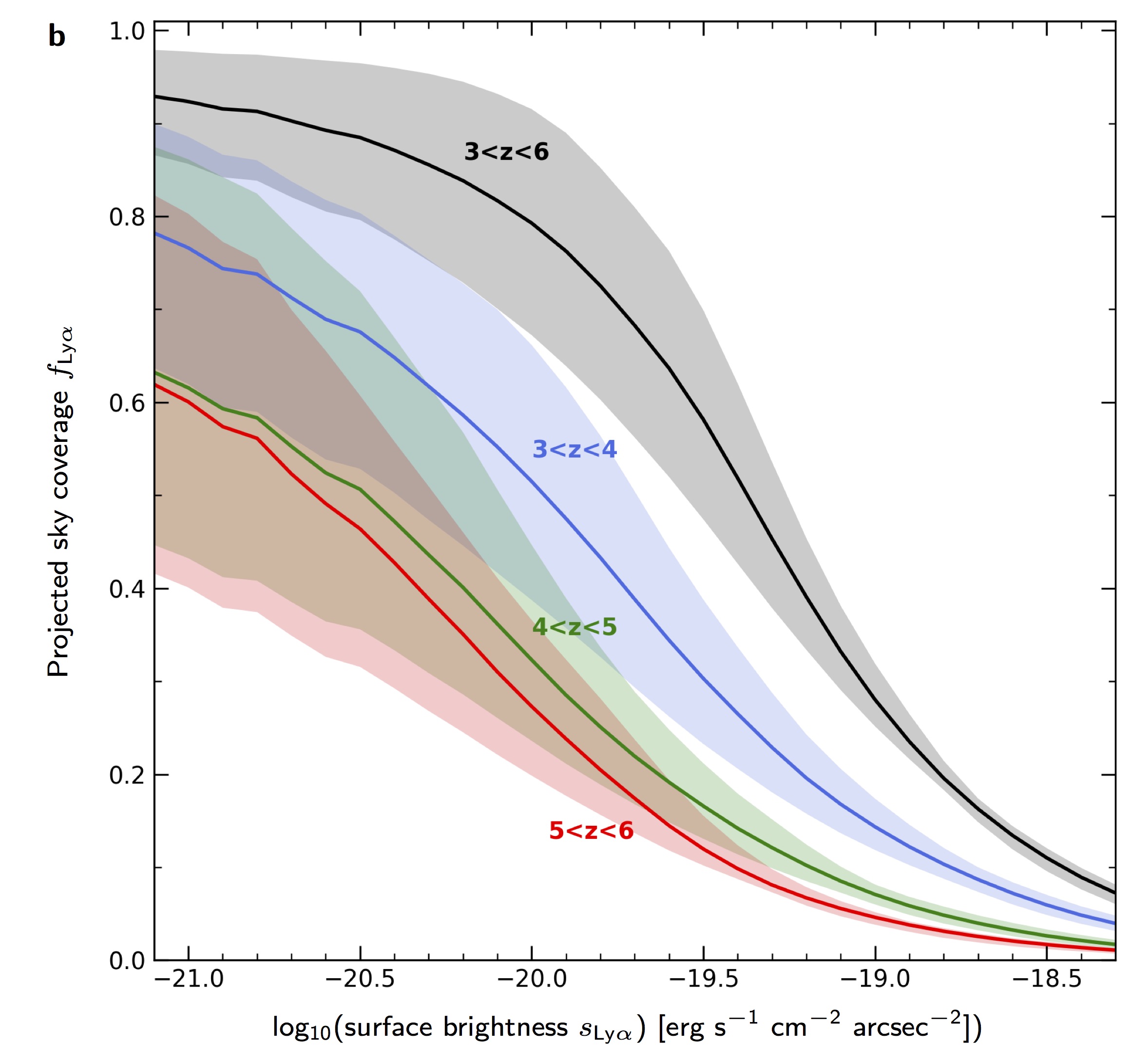}
\caption[]{\textbf{The Lyman-$\alpha$ sky coverage from median-stacked profiles}. \textbf{a,} Reconstructed noise-free Ly$\alpha$ model images of the $1'\times 1'$ section of the HUDF, separately for three disjoint redshift bins marked by the blue/green/red colours, and for the full redshift range shown in black. The shading indicates surface brightness in logarithmic stretch. At the observed position of each object a source model was inserted, rescaled to its actual Ly$\alpha$ flux. The light brown areas delineate the MUSE field of view and mask bright foreground objects. \textbf{b,} The cumulative fractional sky coverage of projected Ly$\alpha$ emission, as a function of limiting surface brightness. Labels and colours indicate the redshift ranges. The solid lines represent the average relations from our two MUSE pointings, while the shaded bands outline our uncertainty estimates from combining the errors of the profiles and the differences between the two fields. This plots demonstrates that the total sky coverage of Ly$\alpha$ emission at $3<z<6$ approaches 100\% for surface brightness levels $s_{\subs{Ly$\alpha$}} \la 10^{-20}$ erg s$^{-1}$ cm$^{-2}$ arcsec$^{-2}$.
}
\end{figure}

\clearpage

\begin{figure}
\includegraphics[height=0.44\textwidth]{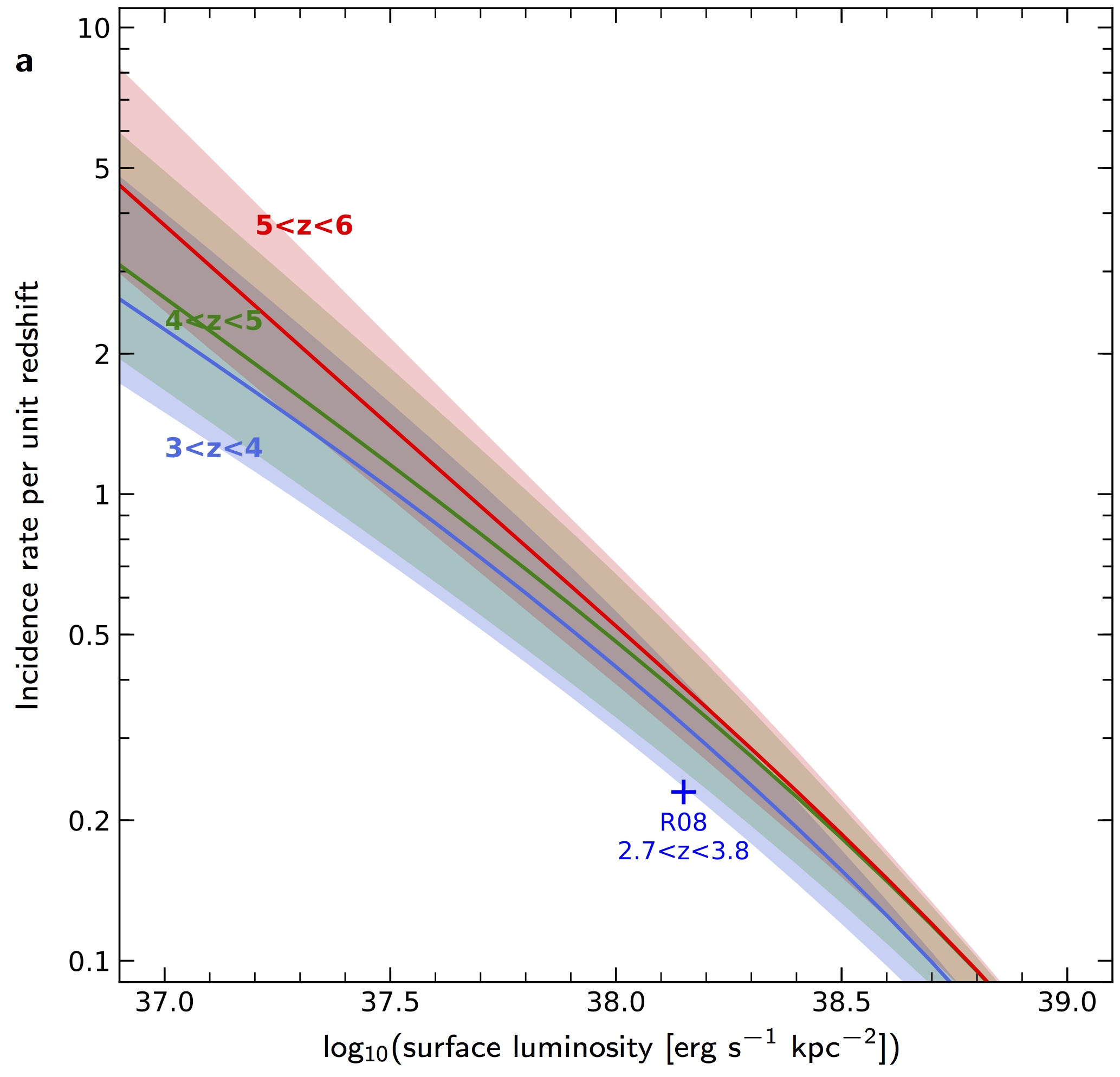}
\hspace{\fill}
\includegraphics[height=0.44\textwidth]{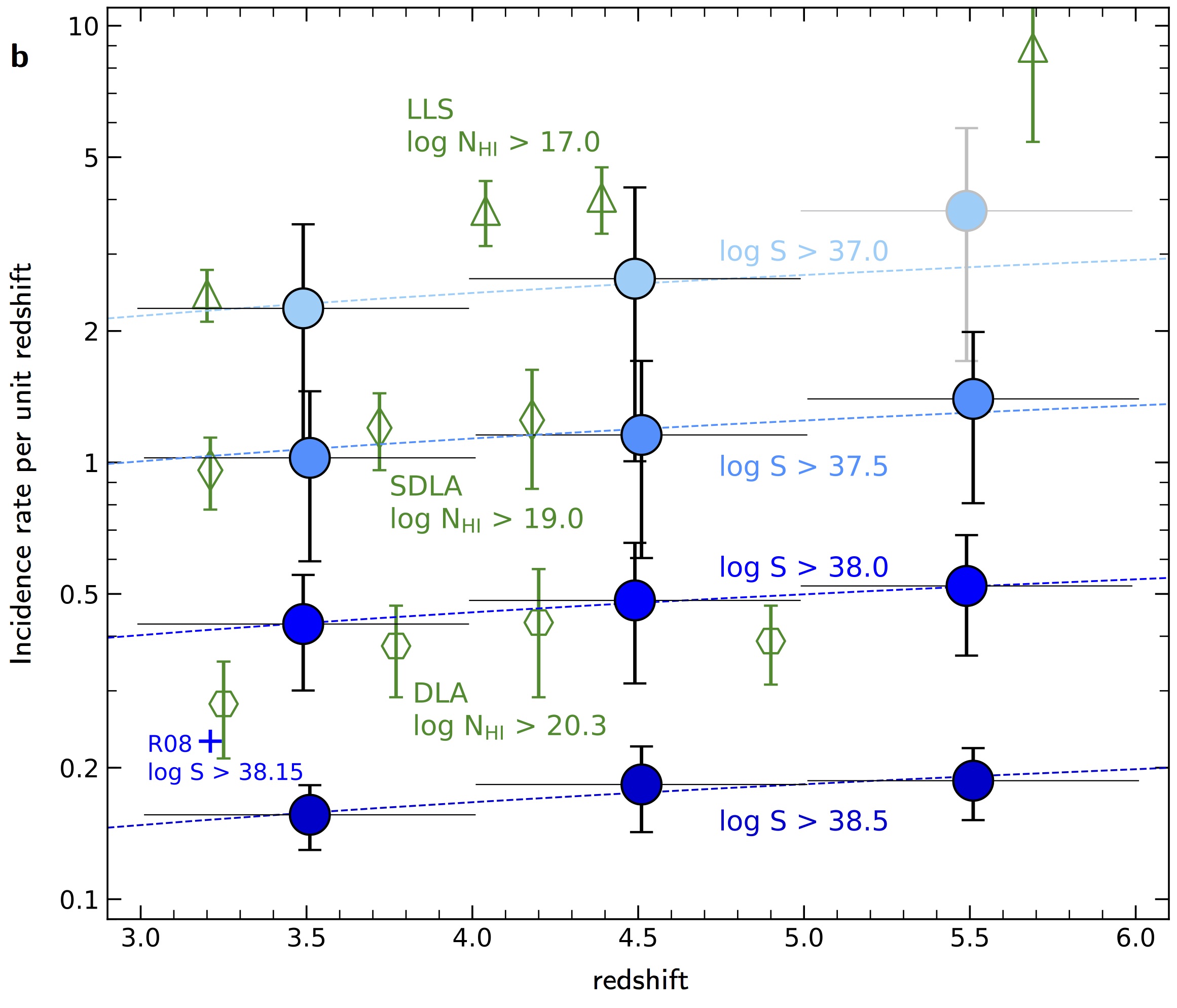}
\caption[]{\textbf{Incidence rates of Ly$\alpha$ emission and comparison with absorption measurements.} \textbf{a,} Cumulative incidence rates of Ly$\alpha$ in emission, as a function of limiting surface luminosity. The shaded bands outline our estimated uncertainties. \textbf{b,} The inferred evolution of Ly$\alpha$ emission incidence rates with redshift, shown as the blue filled circles connected by thin dotted lines. In order to avoid overlapping error bars the symbols are offset by $\pm 0.01$ in redshift. The blue cross in both panels represents the only previous observational estimate of the Ly$\alpha$ emission incidence rate from ref.~\cite{Rauch:2008jy}. Literature values of the cumulative incidence rates of atomic hydrogen measured from quasar absorption lines are provided by the green open symbols, for three commonly adopted limits in column density (triangles: Lyman limit systems\cite{Songaila:2010ft}; diamonds: sub-damped Ly$\alpha$ absorbers\cite{Zafar:2013ew}; hexagons: damped Ly$\alpha$ absorbers\cite{Peroux:2003eg,Crighton:2015jc}). The thin dotted lines show the expected trend for an intrinsically non-evolving population of Ly$\alpha$ emitters.}
\end{figure}

\clearpage

\begin{methods}

\subsection{MUSE observations.}
MUSE is an integral field spectrograph mounted on Unit Telescope 4 of the ESO Very Large Telescope. In its Wide Field Mode it offers a $1'\times 1'$ field of view at a spatial sampling of $0\farcs2\times 0\farcs2$, producing a datacube with 90\,000 spatial pixels. Each spatial pixel contains a 475--935~nm spectrum with $\sim$0.25~nm spectral resolution. The first of the two deep field observations used here was obtained in 2014, where we integrated for a total of 27h on a single pointing in the Hubble Deep Field South\cite{Casertano:2000fn} (HDFS). The data reduction and construction of the first redshift catalogue in this field are summarised in ref.~\cite{Bacon:2015eh}. The second deep MUSE exposure was obtained between 2014 and 2016 as part of a greater effort to perform a contiguous spectroscopic mapping of the Hubble Ultra-Deep Field\cite{Beckwith:2006hp} (HUDF). These observations resulted in a $3'\times 3'$ mosaic with a mean integration time of 10h, on top of which 21h of additional exposure time were dedicated to a single MUSE pointing inside the HUDF. The characteristics of the MUSE-HUDF dataset and the reduction process are described in ref.~\cite{Bacon:2017hn}. Here we use only the two ultra-deep 1~arcmin$^2$ MUSE pointings, which for simplicity we refer to as HUDF and HDFS, respectively. The average spatial resolution (FWHM of the best-fitting Moffat\cite{Moffat:1969ts} function) in the combined coadded datacube, evaluated at 700~nm, is 0\farcs66 for the HDFS and 0\farcs63 for the HUDF.

\subsection{The sample of Ly$\alpha$ emitters. }
Here we focus on galaxies marked by their Ly$\alpha$ emission (Ly$\alpha$ emitters, LAEs). Given the MUSE spectral range of 475--935\,nm, LAEs can be detected over a redshift interval of $2.92 < z < 6.64$.  In order to construct a homogeneous Ly$\alpha$-selected sample we ran our dedicated software LSDCat\cite{Herenz:2017er} (Line Source Detection and Cataloguing) on both datacubes to produce a list of emission line objects, which we then inspected visually to assign redshifts. This resulted in a sample of 128 LAEs in the HDFS and 169 LAEs in the HUDF, respectively. Because of the strictly Ly$\alpha$-based selection, these samples are not mere subsets of our previously published catalogues\cite{Bacon:2015eh,Inami:2017bm} but also contain a few additional LAEs. The spatial distribution of the objects is visualised in Extended Data Fig.~\ref{fig:laesample}. For each LAE, the LSDCat software measures the spatial and spectral centroids and the integrated Ly$\alpha$ fluxes, quantities used in this study. Redshifts were assigned from the measured centroids of the Ly$\alpha$ emission line. While these centroids are known to be shifted with respect to the systemic redshifts by up to a few 100 km/s, accurate knowledge of the latter is not required for the present study (and was not available for our sample). Because of the decrease in instrument sensitivity close to the low and high wavelength cutoffs, and because of the crowding of OH night-sky emission lines towards the reddest wavelengths, we limited our sample to the redshift range $3 < z < 6$, which conveniently allowed us to define three broad redshift bins of $\Delta z = 1$ each. The final sample encompasses 119 LAEs in the HDFS and 151 LAEs in the HUDF, respectively.

\subsection{Extraction of narrowband images.}
Each LAE enters into the current investigation as a pseudo-narrowband (NB) Ly$\alpha$ image, extracted from the datacube at the location of the 3-dimensional Ly$\alpha$ centroid coordinates provided by LSDCat. These images were constructed in the following way: We first extracted a provisional Ly$\alpha$ spectrum from the continuum-subtracted datacube by summing over an unweighted circular aperture of radius 0\farcs6. We then modelled the Ly$\alpha$ emission line profile as a Gaussian, which provided an improved line centroid as well as an approximate line width. Using this Gaussian approximation as spectral template, we performed a weighted summation of spectral layers of the datacube into a single NB image. While Ly$\alpha$ lines usually show some deviations from a single Gaussian, these deviations have negligible effect on the extraction results, except for cases of secondary line peaks (``blue bumps'') which are not captured by our narrowband images. The maximum Gaussian FWHM for the extraction was set to 500 km/s in order to limit the noise. Compared to an unweighted summation over a given bandwidth this scheme provides a better signal-to-noise ratio, and the fractional weights ensure that the bandwidth always matches the actual line width, which matters especially for relatively narrow lines with FWHM $\la$ 200~km/s. We verified that for broader lines the results are very similar to an unweighted summation. The blank-sky noise level in these NB images varies substantially, depending on the spectral bandwidth and on the wavelength of the Ly$\alpha$ line. A typical value of the pixel-to-pixel rms in regions with no detected emission is $5\times 10^{-19}$ erg s$^{-1}$ cm$^{-2}$ pixel$^{-1}$, corresponding to a 1$\sigma$ surface brightness limit of $10^{-19}$ erg s$^{-1}$ cm$^{-2}$ arcsec$^{-2}$ when averaged over an aperture of 1\arcsec. This limit varies by a factor of $\sim$2 between different objects.

\subsection{The Ly$\alpha$ sky coverage from direct projection.}
We obtained a projected Ly$\alpha$ view of each field by coadding all extracted NB images, maintaining the position of each object in the plane of the sky. In order to reduce the noise in the coadded image we introduced a truncation radius of 6\arcsec\ around the centroid of each LAE beyond which the NB data were set to zero for the coadding procedure. This was motivated by the fact that beyond this radius we find generally no individually detectable Ly$\alpha$ emission around our objects. The projection therefore accounts only for the \emph{circumgalactic} Ly$\alpha$ emission around detected sources, and any putative extended Ly$\alpha$ emission at the same redshift as a detected object, but outside the truncation radius, is ignored in the coadding procedure. At the same time the truncation ensures that same-redshift pairs enter only once into the coadded image; if necessary we constructed additional masks by hand to ensure that this was always the case. Masks were also applied in a few cases of contamination by foreground emission lines or by continuum subtraction residuals from bright foreground objects such as stars. Finally, we removed 33 objects where Ly$\alpha$ happened to fall close to a bright sky line causing significant sky-subtraction residuals in the narrowband images. Altogether the amount of budgeted Ly$\alpha$ flux in this approach should be seen as a strict lower limit to the true emission in the field.

With these provisions we estimated the cumulative fractional sky coverage $f_{\subs{Ly$\alpha$}}$ of Ly$\alpha$ emission brighter than $s_{\subs{Ly$\alpha$}}$, documented in Extended Data Fig.~\ref{fig:coadded}. Without spatial filtering the surface brightness is specified for a single pixel of only $0\farcs2\times 0\farcs2$, resulting in extremely noisy images. We therefore filtered the images with a Gaussian of FWHM = 7 pixels (1\farcs4) which provided a good compromise between noise suppression, enforcing large-scale spatial coherence, and ensuring that flux redistribution from the central pixels into the outer parts can be neglected. The maximally reachable covering fraction is limited by random fluctuations due to noise, as can be seen very clearly in Extended Data Fig.~\ref{fig:coadded}: $f_{\subs{Ly$\alpha$}}$ measured from the unfiltered data converges to considerably lower values than $f_{\subs{Ly$\alpha$}}$ measured from the filtered data -- which are of course also affected, but to a lesser degree. We have not attempted to push this approach any further, as for the main results of this paper we employed the stacking-and-insertion modelling approach described below. We did, however, perform a retrospective consistency check between the direct projection and the stacking approach, which worked as follows: We took the idealised noise-free reconstructed model images obtained from the median-stacking analysis (the bottom-right image in Fig.~3a in the case of the HUDF), degraded them by adding realistic noise, and after spatial filtering we measured $f_{\subs{Ly$\alpha$}}$ in the same way as in the real data. For the noise model we filled empty datacubes with normally distributed random numbers scaled to the effective noise in the actual data. From these noise-only cubes we then extracted NB images with the same prescriptions as for the real LAEs, using the same spectral bandpasses and spatial masks, and coadded them to provide a random noise realisation of the projected Ly$\alpha$ image, which we then added to the stacking-based model image. The grey curves in Extended Data Fig.~\ref{fig:coadded} show that these very different approaches to measure $f_{\subs{Ly$\alpha$}}$ produce remarkably consistent results when taking the noise into account. The only noteworthy discrepancy between the thick black and the thick grey lines in Extended Data Fig.~\ref{fig:coadded} occurs around $\log_{10} s_{\subs{Ly$\alpha$}} \sim -18.5$, mainly caused by the median-taking in the stacking process which removes the largest and most extended Ly$\alpha$ emitters. At these relatively high surface brightnesses, the direct projection approach actually delivers a more realistic estimate of $f_{\subs{Ly$\alpha$}}$ than the stacking approach.

\subsection{Stacking analysis and profile modelling.}
We excised $20''\times 20''$ MUSE-narrowband subimages centred on each source and put these into image stacks, separately for the three redshift intervals $3<z<4$, $4<z<5$, and $5<z<6$. Prior to the analysis, the data were subjected to a rigorous visual screening. In this step, 76 objects were removed from the stacks because their NB images were disturbed by sky subtraction residuals or residual emission from unrelated objects. 194 LAEs remained in the combined stacking sample. Bright foreground objects and the edges of the field of view were masked. We also ensured that when the NB image contained multiple objects at the same redshift, each spatial pixel contributed to the stack only once. Finally, each stack was collapsed into a single image by computing the pixel-by-pixel median of all unmasked input image pixels. We chose the median instead of the mean to avoid that possible faint undetected companions enhance the signal in the outskirts. For each collapsed stack we also obtained a weight image containing in each pixel the number of input images that contributed to it after masking; these weight images are shown in Extended Data Fig.~3. 

To detect the low surface brightness Ly$\alpha$ emission in the outer regions of our haloes we extracted azimuthally averaged radial profiles from the median stacks (Fig.~2c), by averaging all pixels within each of a set of concentric annuli (Fig.~2b) defined as follows. Let $r_i$ denote the outer radius of annulus $i$ with $i = 1\dots 11$. For $i\le 5$ we adopted constant annular widths, $r_i = i\times 0\farcs2$ (1 MUSE spatial pixel), for the outer annuli $i\ge 6$ we constructed a progression of increasing widths with the recursion formula $r_i = r_{i-1} + 10^{1.47\times(i-5)/7}\times 0\farcs2$. Thus, the last annulus has an outer radius $r_{11} = 9\farcs97$, just fitting into our $20\arcsec\times 20\arcsec$ images and combining 4660 MUSE spatial pixels into a single surface brightness measurement.

We estimated the uncertainties of these surface brightness profiles in two ways, by formal error propagation and empirically from empty regions in the data. The formal errors originate in the pixel variances of the MUSE datacubes, corrected for resampling effects\cite{Bacon:2017hn}, propagated first into the NB images of individual objects and then into the median-combined stacks. For the latter step we used the property of the median that its variance is approximately $\pi/2$ times the variance of the mean. For the empirical noise calibration we shifted the whole LAE sample in wavelength while maintaining the spatial positions of all objects, thus defining empty regions by applying small redshift offsets. Offsetting in increments of $\pm 5$ MUSE spectral pixels or $\pm 6.25$~\AA\ yielded 40 complete sets of as many empty regions as LAEs, which were then subjected to the same NB image extraction and median stacking procedure as the Ly$\alpha$ data. We estimated the noise from the dispersion of the extracted radial profiles between these 40 sets. The outcome of this experiment is presented in Extended Data Fig.~4. This figure thus provides the significance limits for the detection of very low surface brightness emission in our stacked data.

While the propagated errors can be easily calculated independently of the object content of the datacubes, they do not account for systematics in the background subtraction. On the other hand, the empirically determined errors automatically include such systematics, but are subject to contamination of the supposedly empty regions by unrelated sources and by sky subtraction residuals, and hence they likely overestimate the true errors. In our experiment the median ratios between empirical and propagated errors for the outer profile regions ($r > 1\arcsec$) were (1.02, 1.33, 1.52) for the three redshift ranges, respectively. For this paper we conservatively adopted only the (larger) empirically determined errors for the profiles, and only these are shown in Fig.~2c and Extended Data Fig.~4. The empty regions experiment also revealed that the mean of the empty regions tends to become slightly negative at small radii $r < 2\arcsec$, implying that we may actually underestimate the surface brightnesses in the inner regions by a very small amount. Since this effect is at the $\sim 1$\% level relative to the measured signal at these radii, we decided to neglect it.

We used GALFIT\cite{Peng:2002di} to model the median-stacked images with smooth 2-dimensional Ly$\alpha$ surface brightness distributions. While in previous work\cite{Wisotzki:2016hw,Leclercq:2017cp} we favoured a double exponential model with a compact core and an extended halo, that model was driven mainly by the high S/N and high surface brightness regions within $\la$10~kpc. In the present study the emphasis is on the outer regions, where Fig.~2 suggests that the surface brightness distribution of the halo may show some flattening relative to a single exponential. Here we adopted a model consisting of a central point source plus a circular Sersic\cite{Sersic:1968ta} function, convolved with the point-spread function at the relevant wavelengths, which provided good fits to all median-stacked images. To quantify the uncertainties in the fitted profiles we used the formal error estimates of GALFIT, but then increased these until the uncertainties of the fits were consistent with the empirically calibrated error bars of the directly extracted azimuthal profiles. These uncertainties are displayed as the shaded regions around the fitted profiles in Fig.~2. Extended Data Table~1 provides the numerical values of the fit parameters and their uncertainty estimates. 

We make the central assumption that the shapes of the Ly$\alpha$ haloes of LAEs do not, in the statistical average, depend on their luminosities. While in ref.~\cite{Leclercq:2017cp} we found no evidence for such a dependence, most of those objects have Ly$\alpha$-luminosities $L_{\mathrm{Ly}\alpha} > 10^{42}$ erg~s$^{-1}$, whereas our current sample has a median $\log_{10} L_{\mathrm{Ly}\alpha}/\mathrm{erg}\:\mathrm{s}^{-1}$ of only 41.7. Here we use the median stacks to probe deeper into the validity of the above assumption. In Extended Data Fig.~5 we present a comparison between the radial profiles extracted from the median stack of the full sample and of a subset with Ly$\alpha$ luminosities $L_{\mathrm{Ly}\alpha} > 10^{42}$ erg~s$^{-1}$. There are (18, 13, 6) objects at $z$ = (3--4,\,4--5,\,5--6) meeting this criterion. The median-stack profiles of the luminous subsets resemble those of the full sample remarkably well. This comparison supports the validity of our self-similarity approximation across the luminosity range of our sample. We plan to revisit this point and related aspects in a follow-up study of Ly$\alpha$ halo profile shapes in stacked MUSE data.

We used the analytic profile fits as templates to construct idealised representations of all LAEs in the two fields, including those objects previously removed from the stacking subsample. All model LAEs at a given redshift range were assigned to have the same spatial surface brightness distribution, but rescaled to the actually observed Ly$\alpha$ fluxes of each real object. Furthermore, the GALFIT models were approximately corrected for PSF blurring by using a delta function (in fact a very narrow Gaussian) as PSF when reconstructing the 2-dimensional model templates. Since for each object the template was rescaled to match the measured Ly$\alpha$ flux, this implies that the modelling of the brightest LAEs involved a certain degree of extrapolation of the profiles. We demonstrate below that the contribution of extrapolated emission to the incidence rates is small (see also Extended Data Fig.~7).

\subsection{Determination of incidence rates.}
We estimated the Ly$\alpha$ emission incidence rates directly from the LAE samples as the sum of circular cross-sections $\pi r^2_\mathrm{iso, i}$ where $r_\mathrm{iso, i}(s_{\subs{Ly$\alpha$}})$ is the isophotal extent of object $i$ at a given surface brightness level, obtained from the GALFIT templates and rescaled to the observed Ly$\alpha$ fluxes. The incidence rate is then
\begin{equation}
   \frac{\mathrm{d}n}{\mathrm{d}z}(s_{\subs{Ly$\alpha$}}) \approx \frac{1}{A_{\subs{FoV}}}\,\frac{n_{\subs{obj}}}{\sum\Delta z_i}\,\sum\limits_{i=1}^{n_{\subs{obj}}} \pi r^2_\mathrm{iso, i}(s_{\subs{Ly$\alpha$}})
   \label{eq:dndz-sample}
\end{equation}
where $A_{\subs{FoV}}$ is the area of the field of view, $\Delta z_i$ is the redshift path length over which object $i$ would be part of the flux-limited sample, and where the summation is carried out over all objects in the redshift range. The normalisation of \dndz\ to a quantity per unit redshift was in our case conveniently provided by the widths of our adopted redshift intervals.

Since $s_{\subs{Ly$\alpha$}}$ decreases with increasing redshift as $(1+z)^4$, we decided to move to distance-independent surface luminosities $S_{\subs{Ly$\alpha$}}$. While this quantity is not much used in the literature, we prefer working with intrinsic object properties over rescaling observed quantities to some fiducial reference redshift. The transformation is $\log_{10} S_{\subs{Ly$\alpha$}} = \log_{10} s_{\subs{Ly$\alpha$}} + 4 \log_{10}(1+z) + 54.71$ when both $s$ and $S$ are given in cgs units. The right-hand ordinate of Extended Data Fig.~\ref{fig:prof} provides a quick-look visual calibration of the conversion.

\subsection{Correction for sample incompleteness.}
Our LAE sample certainly suffers from incompleteness close to the flux limit, with faint objects getting selected only if their Ly$\alpha$ emission is sufficiently pointlike. Selection effects for the detection of LAEs in the MUSE-HUDF survey were investigated in detail in ref.~\cite{Drake:2017jq}, where it was shown that the transition from 80\% to 20\% detection probability extends over $\sim$0.5~dex in line flux. The Ly$\alpha$ incidence rates calculated from Eq.~\ref{eq:dndz-sample} are therefore biased low, missing the contributions from undetected but presumably existing objects. In order to correct for this incompleteness we replaced the summation over the observed sample by an integration over the full survey volume, assuming that the intrinsic distribution of Ly$\alpha$ luminosities follows the luminosity function determined in ref.~\cite{Drake:2017jq}. Since the self-similarity approximation of the extended Ly$\alpha$ emission implies a unique relation between the total flux $F_{\subs{Ly$\alpha$}}$ of an object and its isophotal radius, $r_\mathrm{iso} = r_\mathrm{iso}(F_{\subs{Ly$\alpha$}}, s_{\subs{Ly$\alpha$}})$, we can predict the emission cross-sections from only the Ly$\alpha$ luminosities. The total incidence rate \dndz\ for a given redshift range ($z_1$, $z_2$) follows as
\begin{equation}
   \frac{\mathrm{d}n}{\mathrm{d}z}(s_{\subs{Ly$\alpha$}}) = \frac{1}{(z_2 - z_1)\,A_{\subs{FoV}}}\,\int\limits_{z_1}^{\;z_2} \mathrm{d}V_\mathrm{c}(z)\int\limits_{\ell_\mathrm{min}}^{\ell_\mathrm{max}} \mathrm{d}\ell \: \pi r^2_\mathrm{iso}[s_{\subs{Ly$\alpha$}}, F_{\subs{Ly$\alpha$}}(\ell,z)] \:\phi(\ell,z)
   \label{eq:dndz-lf}
\end{equation}
where $V_\mathrm{c}(z)$ is the differential comoving volume element at redshift $z$, $\ell \equiv \log_{10}(L_{\subs{Ly$\alpha$}}) = \log_{10}(4\pi d_\mathrm{L}^2\,F_{\subs{Ly$\alpha$}})$ (in units of erg s$^{-1}$, with the cosmological luminosity distance $d_\mathrm{L}$), and $\phi(\ell,z)\,\mathrm{d}\ell$ is the differential Ly$\alpha$ luminosity function at redshift $z$. We parametrised $\phi(\ell)$ as a redshift-independent Schechter function with $\ell^\star = 42.59$, $\phi^\star = 2.138\times 10^{-3}$ Mpc$^{-3}$, and $\alpha = -1.93$, which is a good overall fit to the completeness-corrected LAE sample\cite{Drake:2017jq}. While the upper luminosity integration limit $\ell_\mathrm{max}$ does not matter much as $\phi(\ell)$ approaches zero very quickly for increasing $\ell$, choosing a value for the lower integration limit $\ell_\mathrm{min}$ is less straightforward: Although faint LAEs have small isophotal radii, they are also numerous and thus contribute non-negligibly to the integrated cross-section. The integral converges only for $\ell_\mathrm{min} \la 40.5$, but at the expense of including large numbers of hypothetical ultrafaint LAEs into the budget that are well below the current detection limits. As our ``best guess'' we adopted $\ell_\mathrm{min} = 41.0$ which is slightly brighter than the faintest detected LAEs in our actual sample. Extended Data Fig.~\ref{fig:dndz} provides a synopsis of the different approaches to estimate \dndz\ from our data. These plots also show that the magnitude of the completeness correction is by far the dominant source of uncertainty for \dndz. We therefore adopted as the lowest reasonable limit the values of \dndz\ without any completeness correction (i.e.\ from Eq.~\ref{eq:dndz-sample}) obtained from the somewhat `emptier' HDFS. As an upper limit we took the asymptotic result from integrating Eq.~(\ref{eq:dndz-lf}) with $\ell_\mathrm{min} = 40.0$. For the presentation in Fig.~4 we interpreted these lower and upper bounds as $\pm 2\sigma$ limits, but plotted only the 1$\sigma$ error envelopes in accordance with the usual conventions.

\subsection{Discussion of systematic errors.}
We first consider to what extent the integrated values of \dndz\ depend on extrapolations of the rescaled Ly$\alpha$ profiles beyond the radial range over which they were constructed. To address this question we constructed the cumulative distribution of the contributions to the total incidence rate sum or integral as a function of their isophotal radii. The results of the completeness-corrected integration of \dndz\ (Eq.~\ref{eq:dndz-lf}) with $\ell_\mathrm{min} = 41.0$ (our `best guess' approach) are shown in Extended Data Fig.~7. These plots demonstrate that at all redshifts and for all levels of $S_{\subs{Ly$\alpha$}}$ except the lowest, more than 80\% of the total incidence rates is contributed by emission from $r_\mathrm{iso} < 5\arcsec$. Only for $\log_{10} S_{\subs{Ly$\alpha$,lim}}/\mathrm{erg}\: \mathrm{s}^{-1}\:\mathrm{kpc}^{-2} = 37$ and $5<z<6$ there is a substantial extrapolated contribution, which is why we consider this point as uncertain and plotted it in light grey in Fig.~4b.

A strong assumption made in our analysis is the azimuthal symmetry of the Ly$\alpha$ emission, which is certainly not strictly correct. We now quantify the biases arising from the circularisation of a non-axisymmetric signal through median stacking. Extended Data Fig.~8a--c shows how an elliptical 2-dimensional surface brightness distribution and the corresponding isophotal cross-sections are modified when the cross-sections are estimated from an azimuthally averaged profile. The circularised profile is broadened and the cross-sections at given surface brightnesses are overestimated, but the effect is quite small. More relevant for our analysis, Extended Data Fig.~8d--f demonstrates that median stacking of randomly oriented elongated objects delivers isophotal cross-sections that are systematically \emph{smaller} than the true values. While the above calculations are based on a rather simple source model, the conclusion can be qualitatively generalised to other non-axisymmetric surface brightness distributions. The median stacking ensures that the derived emission cross-sections, and consequently the inferred Ly$\alpha$ sky coverage and incidence rates, are rather under- than overestimated.

While the small scale distribution of Ly$\alpha$ emission remains unknown at these surface brightness levels, we have some idea of the projected covering fraction $f_{\mbox{\scriptsize\ion{H}{i}}}$ of neutral hydrogen close to galaxies. Ref.~\cite{Rahmati:2015jf} used the cosmological EAGLE simulation\cite{Schaye:2015gk} to show that $f_{\mbox{\scriptsize\ion{H}{i}}}$ depends on many parameters: distance to galaxy centre, column density, redshift, halo mass, environment. \ion{H}{i} absorption line measurements close to galaxies at $z\simeq 2.5$\cite{Rudie:2012is} give $f_{\mbox{\scriptsize\ion{H}{i}}}=0.3\pm0.14$ for Lyman Limit Systems within one virial radius ($r_\mathrm{vir}$); there are so far no good measurements for $z>3$. Simulations predict that $f_{\mbox{\scriptsize \ion{H}{i}}}$ increases rapidly towards higher redshift\cite{Rahmati:2015jf}, and we expect $f_{\mbox{\scriptsize\ion{H}{i}}}\sim 0.4$--0.8 for $r < r_\mathrm{vir}$ at the redshifts of our sample. The virial radii of our LAEs are unknown, but predicted to be around 30~kpc or less\cite{Garel:2015fi}. Consulting again Extended Data Fig.~7 we see that except for $\log_{10} S_{\subs{Ly$\alpha$,lim}}/\mathrm{erg}\: \mathrm{s}^{-1}\:\mathrm{kpc}^{-2} = 37$ and $5<z<6$, more than 80\% of the integrated Ly$\alpha$ emission incidence rates come from radii less than 30~kpc, i.e.\ from within one virial radius. Unless the Ly$\alpha$-emitting gas has a very different spatial distribution than the general circumgalactic \ion{H}{i}, the covering fractions $f_{\mbox{\scriptsize\ion{H}{i}}}$ are expected to be sufficiently close to unity that the systematic errors from any non-axisymmetry of our derived incidence rates should be small.

A rather different systematic error arises from the limitation of our sample to galaxies selected by their Ly$\alpha$ emission. If not all galaxies are LAEs, then there is circumgalactic \ion{H}{i} gas contributing to high-column density absorption systems which is not included in our budget of extended Ly$\alpha$ emission. The fraction of galaxies at $z>3$ showing detectable Ly$\alpha$ emission depends strongly on the selection criteria. While only some 10\%--20\% of continuum-bright galaxies at these redshifts are strong LAEs with Ly$\alpha$ rest-frame equivalent widths (EW) greater than 50~\AA\cite{Stark:2010hv}, this fraction probably increases to $\sim$50\% if also weaker emitters are included\cite{Steidel:2011jk,Caruana:2018jy}. There are indications that the fraction of strong emitters may be even considerably larger than 50\% for very low luminosity galaxies and/or higher redshifts\cite{Stark:2010hv,Stark:2011bh}. It seems thus plausible to estimate that our Ly$\alpha$ selection captured roughly half of all galaxies at these redshifts. If the other 50\% have a circumgalactic medium similar to the LAEs except for the lack of Ly$\alpha$ emission -- this is a very uncertain assumption --, the incidence rates of the circumgalactic absorbers in Fig.~4 should be shifted downwards by $\sim$0.3~dex to account only for the LAE fraction. Interpolating between the measured values, a surface luminosity level of $\log_{10} S_{\subs{Ly$\alpha$}}/\mathrm{erg}\: \mathrm{s}^{-1}\:\mathrm{kpc}^{-2} \simeq 37.5$ would then have roughly the same incidence rate as absorbers with $\log_{10} N(\mbox{\ion{H}{i}})/\mathrm{cm}^{-2} \sim 18$, less than the limit for SDLAs but still optically thick to Lyman continuum radiation. On the other hand, there may also be non-LAE galaxies with still undetected faint Ly$\alpha$ haloes, similar to those found in ref.~\cite{Steidel:2011jk}, which would increase the Ly$\alpha$ incidence rates even further.

\subsection{Data availability. }
The observations of the HUDF discussed in this paper were made using European Southern Observatory (ESO) Telescopes at the La Silla Paranal Observatory under programme IDs 094.A-0289, 095.A-0010, 096.A-0045 and 096.A-0045. The corresponding data are available on the ESO archive at http://archive.eso.org/cms.html. The data of the HDFS were obtained during MUSE commissioning observations and are available at http://muse-vlt.eu/science/hdfs-v1-0/.

\end{methods}

\clearpage
\section*{Additional references}

\clearpage

\renewcommand{\figurename}{\textbf{Extended Data Fig.}}
\renewcommand{\tablename}{\textbf{Extended Data Table}}

\setcounter{figure}{0}

\begin{figure}
\includegraphics[width=\textwidth]{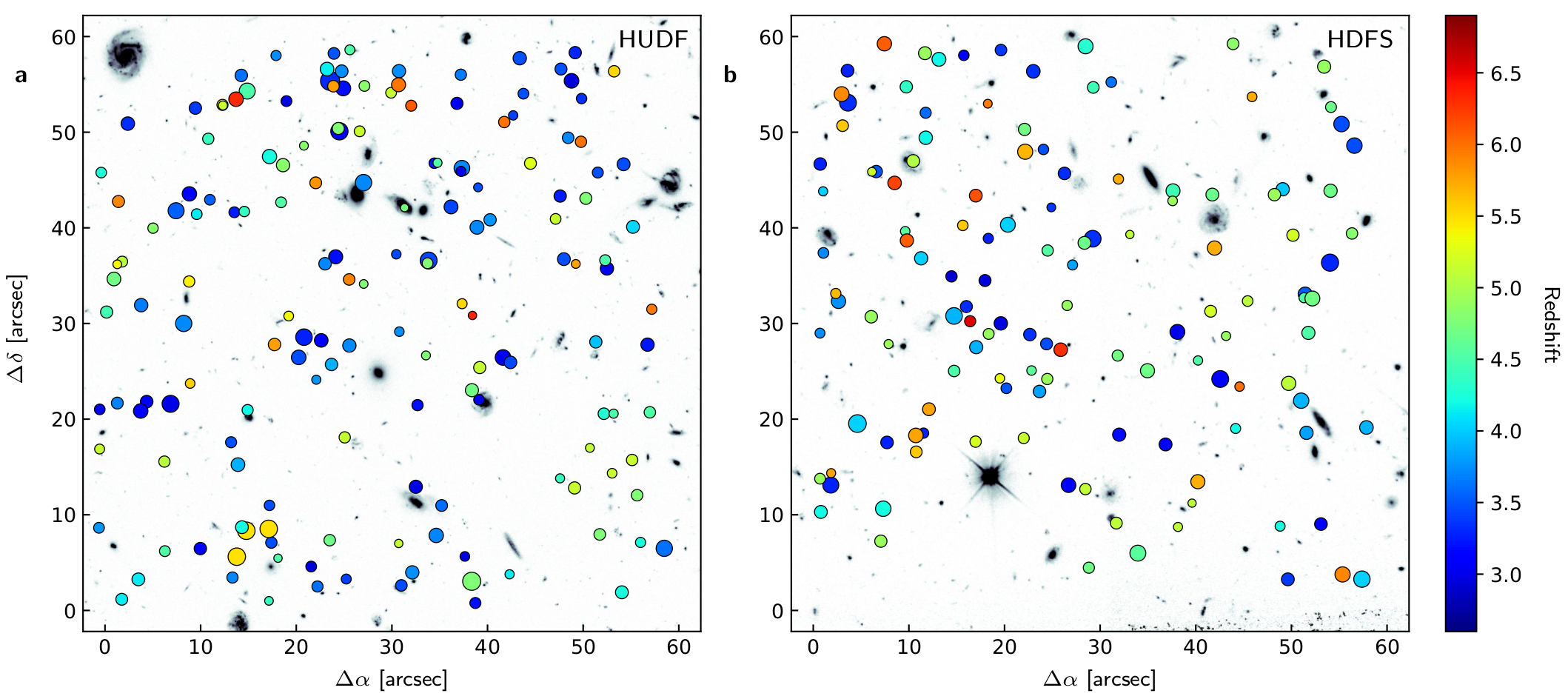}
\caption[]{\textbf{Spatial distribution and redshifts of the Ly$\alpha$ emitter sample.} \textbf{a,} The region observed with MUSE in the Hubble Ultra Deep Field, \textbf{b,} the same for the Hubble Deep Field South. Each Ly$\alpha$ emitter is represented by a circle colour-coded by redshift and with a radius scaled by the integrated Ly$\alpha$ flux of the object. There are several cases of significant crowding of unequal-redshift objects separated by less than a few arcseconds in projection. The underlying greyscale images show the two fields as seen with HST. 
}
\label{fig:laesample}
\end{figure}

\begin{figure}
\includegraphics[width=\textwidth]{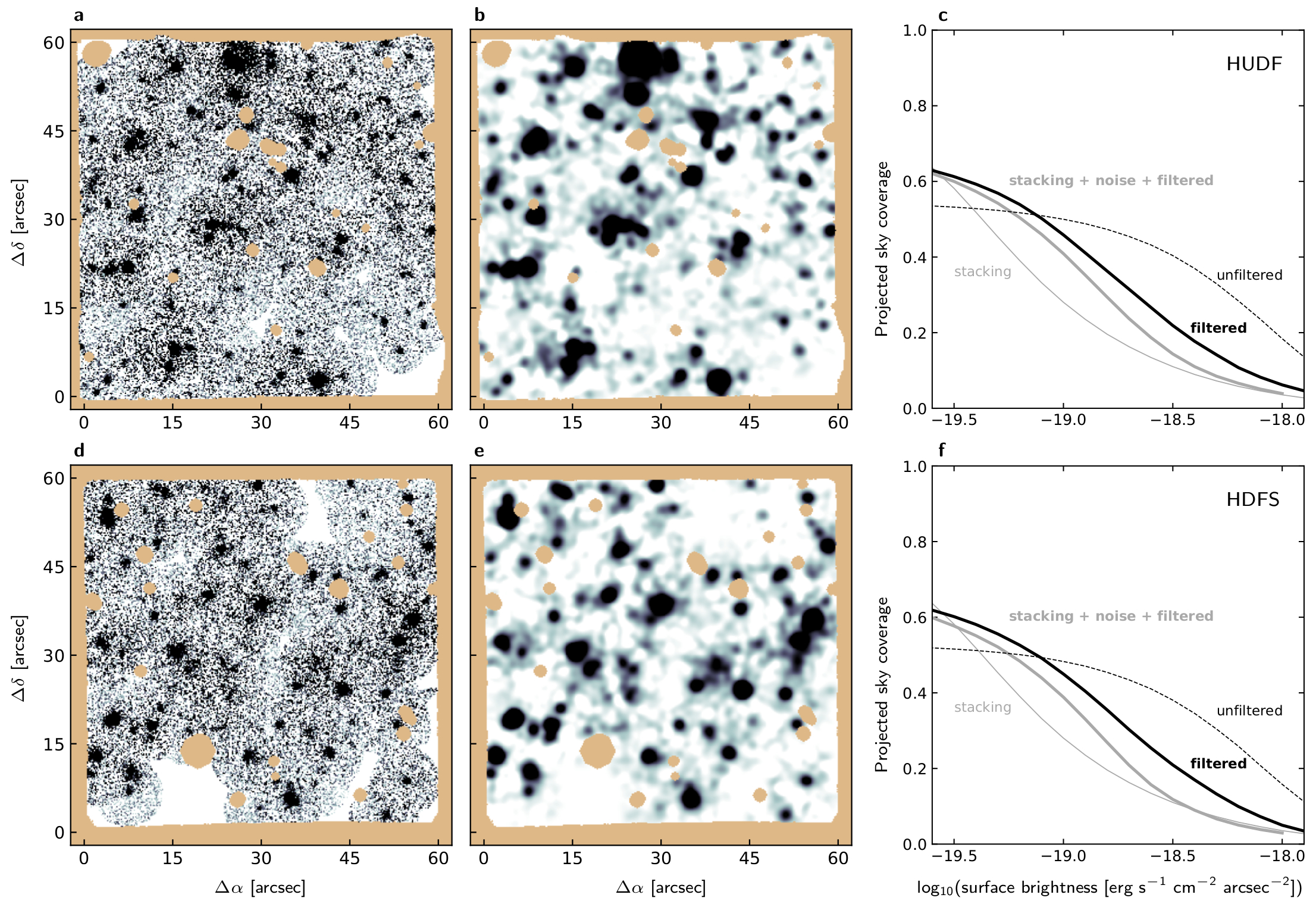}
\caption[]{\textbf{Ly$\alpha$ sky coverage from direct projection.}  
The top row presents the HUDF, the bottom row the HDFS. The greyscale images display the projected and coadded Ly$\alpha$ emission over the redshift range $3<z<6$: left without any spatial filtering, and in the middle after Gaussian filtering with FWHM = 1\farcs4. The filtered image of the HUDF in the top middle is the same as the blue overlay in Fig.~1. The light brown areas delineate the MUSE field of view and indicate masked bright foreground objects. The right-hand panels show the resulting fractional sky coverage, the black dashed line representing the unfiltered and the thick black solid line representing the spatially filtered images. The thin grey line provides the result from the stacking analysis (Fig.~3) for comparison, and the thick grey line is a fiducial reconstruction of the sky coverage derived from a noisy and filtered version of the stacking-based model images. See the discussion in the Methods section for an interpretation of these figures.
}
\label{fig:coadded}
\end{figure}

\begin{figure}
\includegraphics[width=\textwidth]{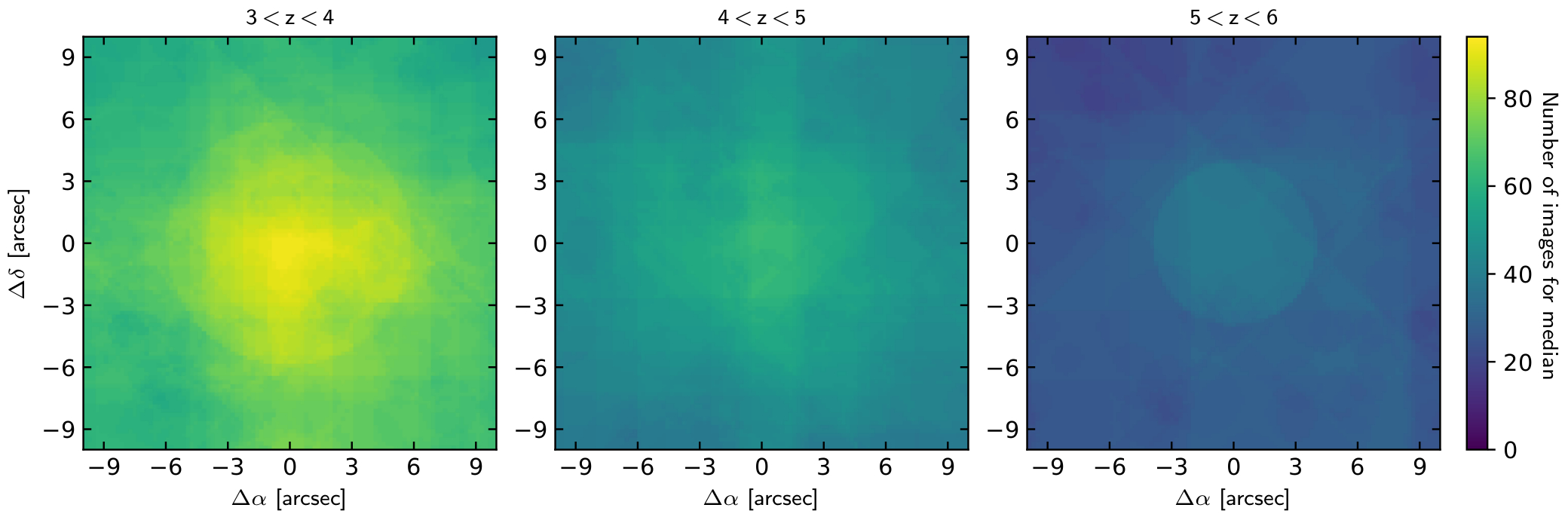}
\caption[]{\textbf{Number of images contributing to the median stacks.} The colour code represents, for each pixel in a median-stacked image, the number of original image pixels contributing to it. This number differs from the total number of objects in a given stack because of the masks applied to several of the contributing images.}
\label{fig:weights}
\end{figure}

\begin{figure}
\includegraphics[width=0.7\textwidth]{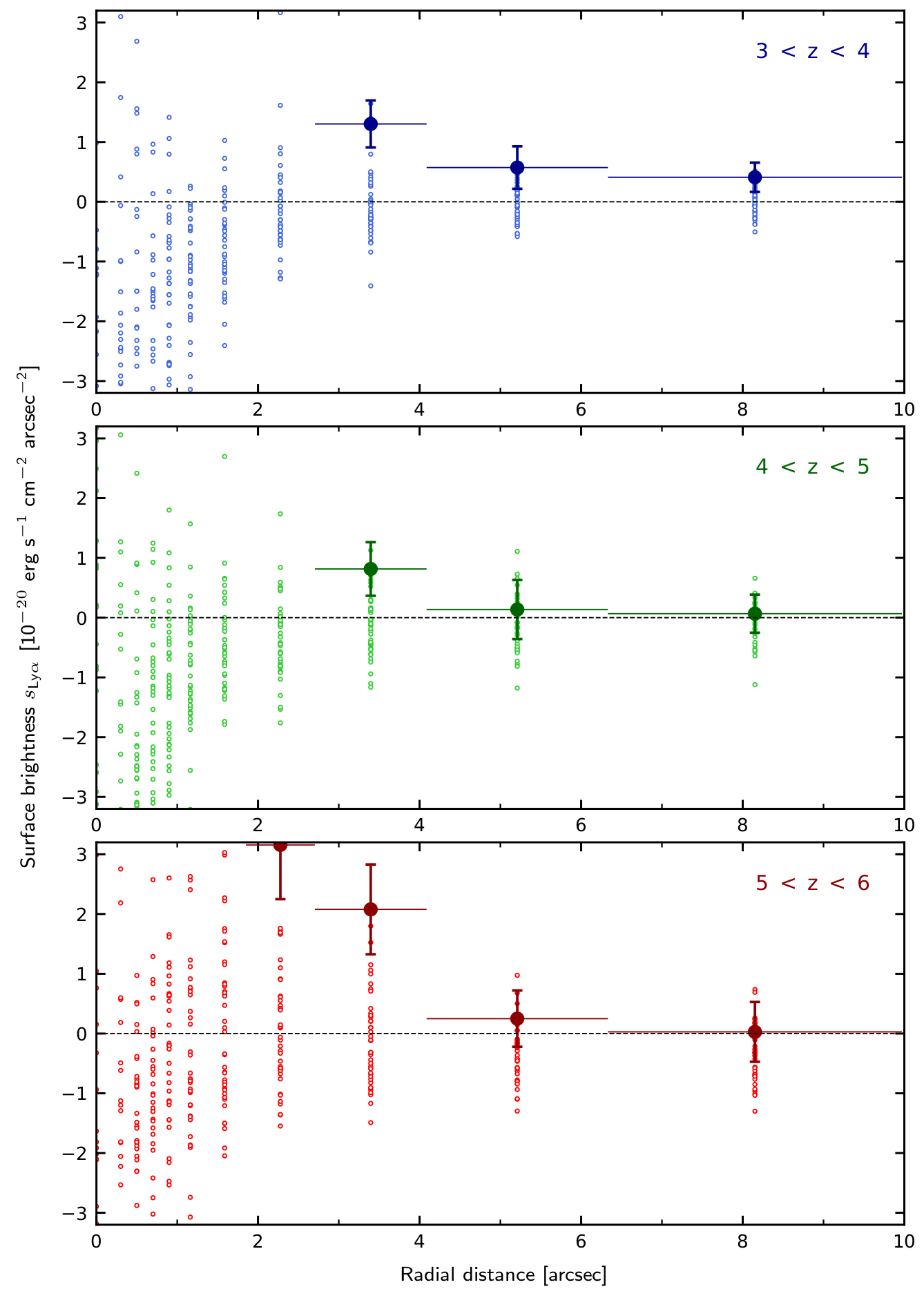}
\caption[]{\textbf{Error calibration and sensitivity of azimuthally averaged profiles.} This figure provides a `zoom view' into the profiles at very low surface brightnesses, with linear ordinate scaling so that negative measurements can be displayed. The small open circles represent the surface brightnesses measured in 40 realisations of a median-stacking analysis of empty regions as described in the Methods section. The filled symbols reproduce the data points from Fig.~2c. The horizontal bars again indicate the widths of the annuli, and the vertical error bars are based on the 1$\sigma$ scatter of the empty field median-stack profiles; these were also adopted as error bars for the data points in Fig.~2c.}
\label{fig:errors}
\end{figure}

\begin{figure}
\includegraphics[width=0.7\textwidth]{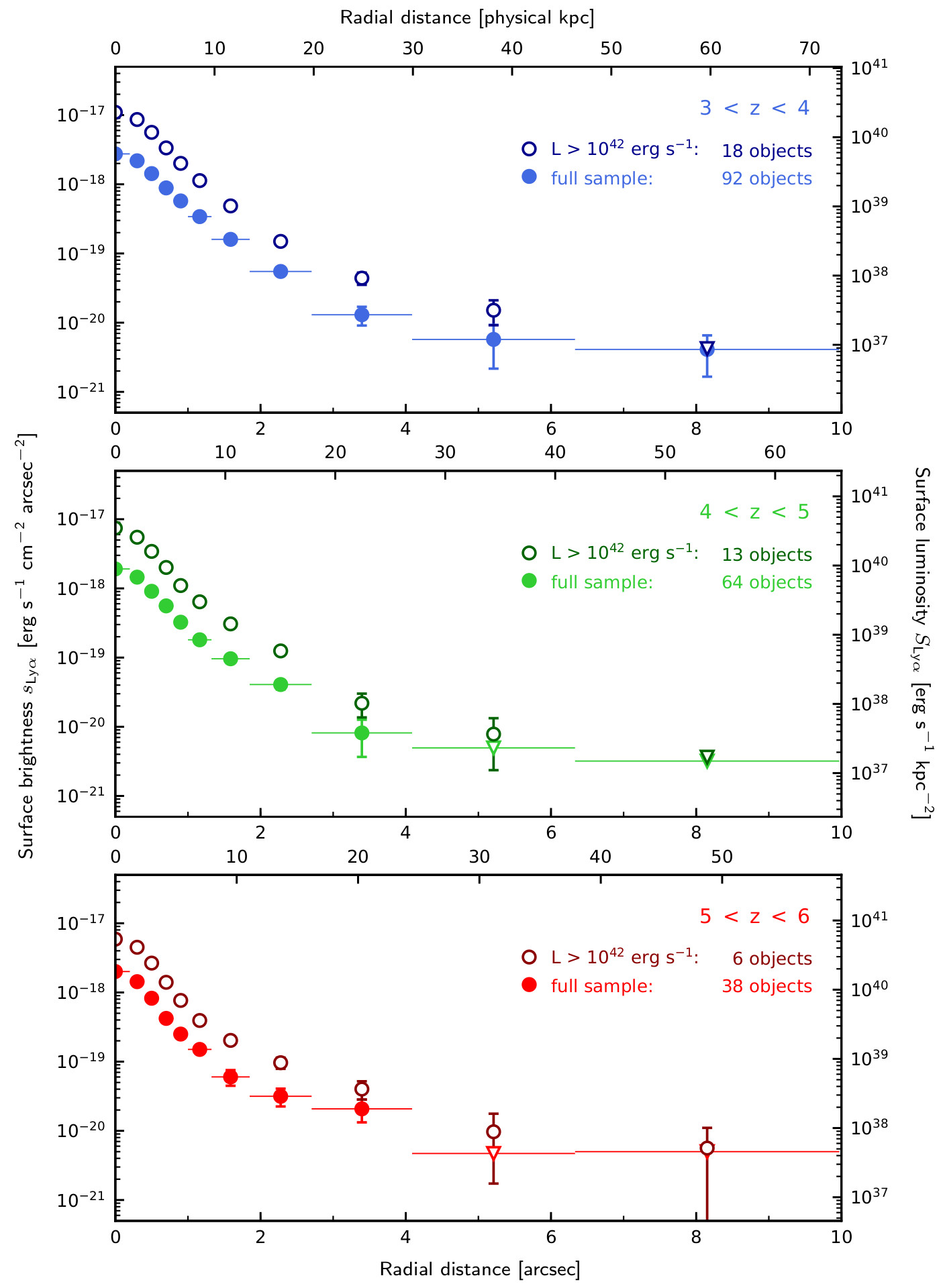}
\caption[]{\textbf{A test of the self-similarity assumption for different Ly$\alpha$ luminosities.} Comparison of azimuthally averaged radial profiles of median-stacked Ly$\alpha$ images above a minimum Ly$\alpha$ luminosity (open dark circles) and with no such cut (light filled circles), for three redshift ranges. As in Fig.~2, the vertical bars on the data points quantify the $1\sigma$ surface brightness measurement errors, while the horizontal bars (drawn only for the filled symbols) indicate the widths of the annuli. Inverted triangles indicate upper limits. The right-hand ordinate provides the conversion from apparent surface brightnesses to redshift-corrected surface luminosities, evaluated at the central redshift of each bin.}
\label{fig:prof}
\end{figure}

\begin{figure}
\includegraphics[width=\textwidth]{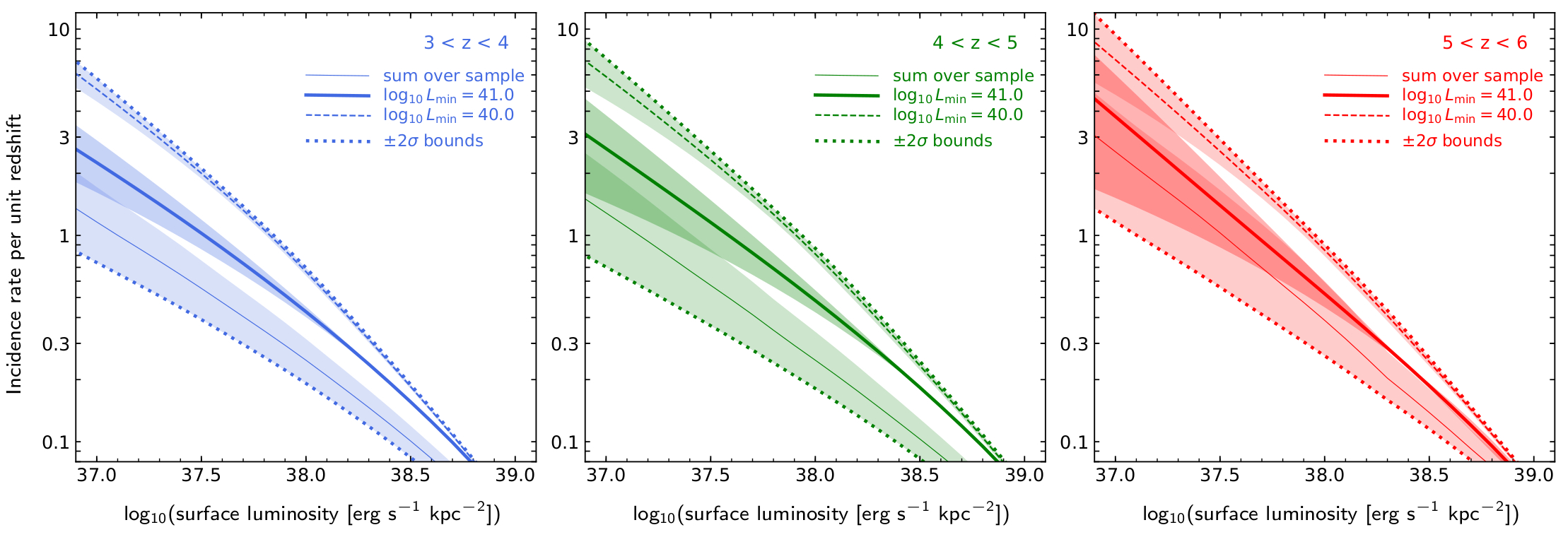}
\caption[]{\textbf{Comparison of approaches to determine the Ly$\alpha$ emission incidence rates.} Each panel shows the cumulative incidence rate as a function of limiting surface luminosity for the specified redshift range, estimated by different methods: Direct summation of Ly$\alpha$ cross-sections over the sample without correcting for incompleteness (Eq.~\ref{eq:dndz-sample}, thin lines), and integrating over the completeness-corrected luminosity function following Eq.~(\ref{eq:dndz-lf}), using lower integration limits of $\ell_\mathrm{min} = 41.0$ (best guess, thick solid line) and $\ell_\mathrm{min} = 40.0$ (asymptotic case, dashed line), respectively. The shaded error bands for direct summation are dominated by field-to-field variance between the HUDF and the HDFS, with the upper envelope tracing the HUDF and the lower envelope tracing the HDFS results. For the luminosity function integration the error bands on these curves incorporate only the statistical uncertainties of the median-stacked profiles. The two thick dotted lines indicate the finally adopted lower and upper $2\sigma$ bounds on the `best guess' results shown in Fig.~4.
}
\label{fig:dndz}
\end{figure}

\begin{figure}
\includegraphics[width=\textwidth]{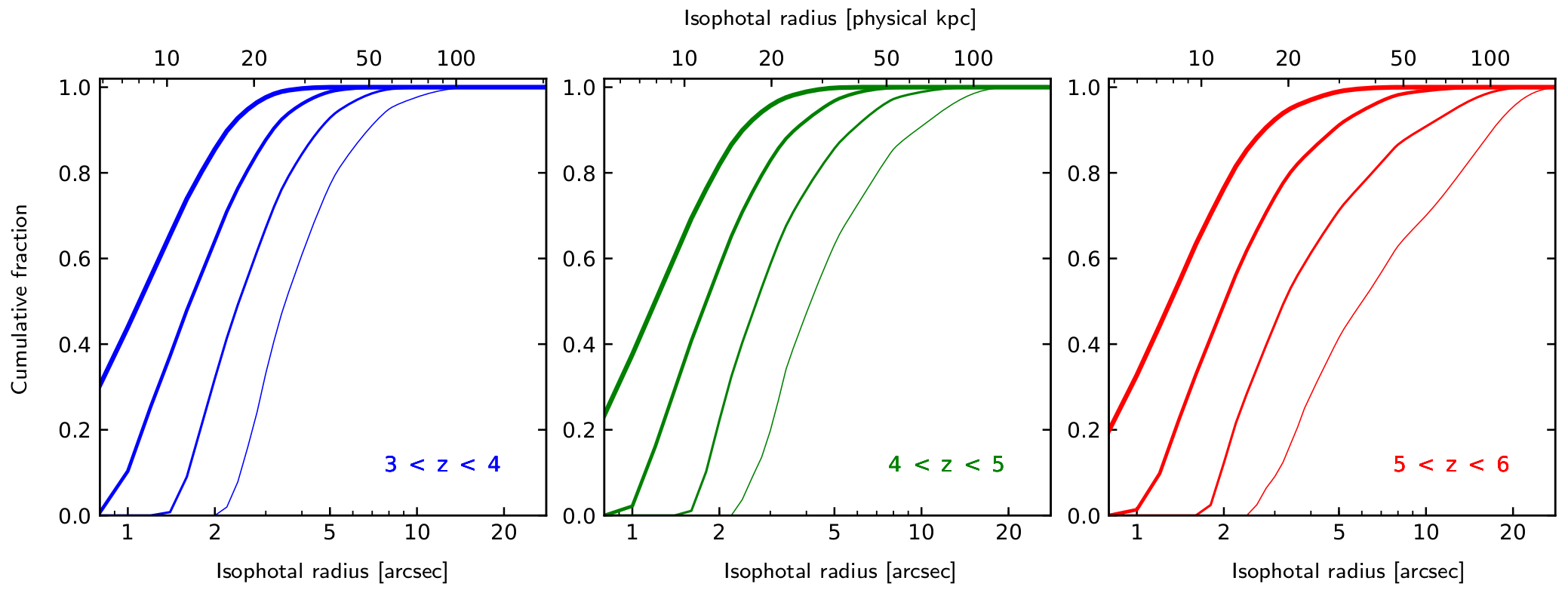}
\caption[]{\textbf{Cumulative contributions to incidence rates as function of radius.} Each panel shows the fractional contributions of objects with different isophotal radii to the integrated Ly$\alpha$ emission cross-section \dndz, using Eq.~\ref{eq:dndz-lf} with $\ell_\mathrm{min} = 41$. The four lines represent, from right to left and with decreasing line width, surface luminosity limits $\log_{10} S_{\subs{Ly$\alpha$}} = 37$, 37.5, 38, 38.5. These plots demonstrate that the Ly$\alpha$ incidence rates are dominated by objects with sizes of a few arcsec (a few tens of kpc).
}
\label{fig:cum}
\end{figure}

\begin{figure}
\includegraphics[width=\textwidth]{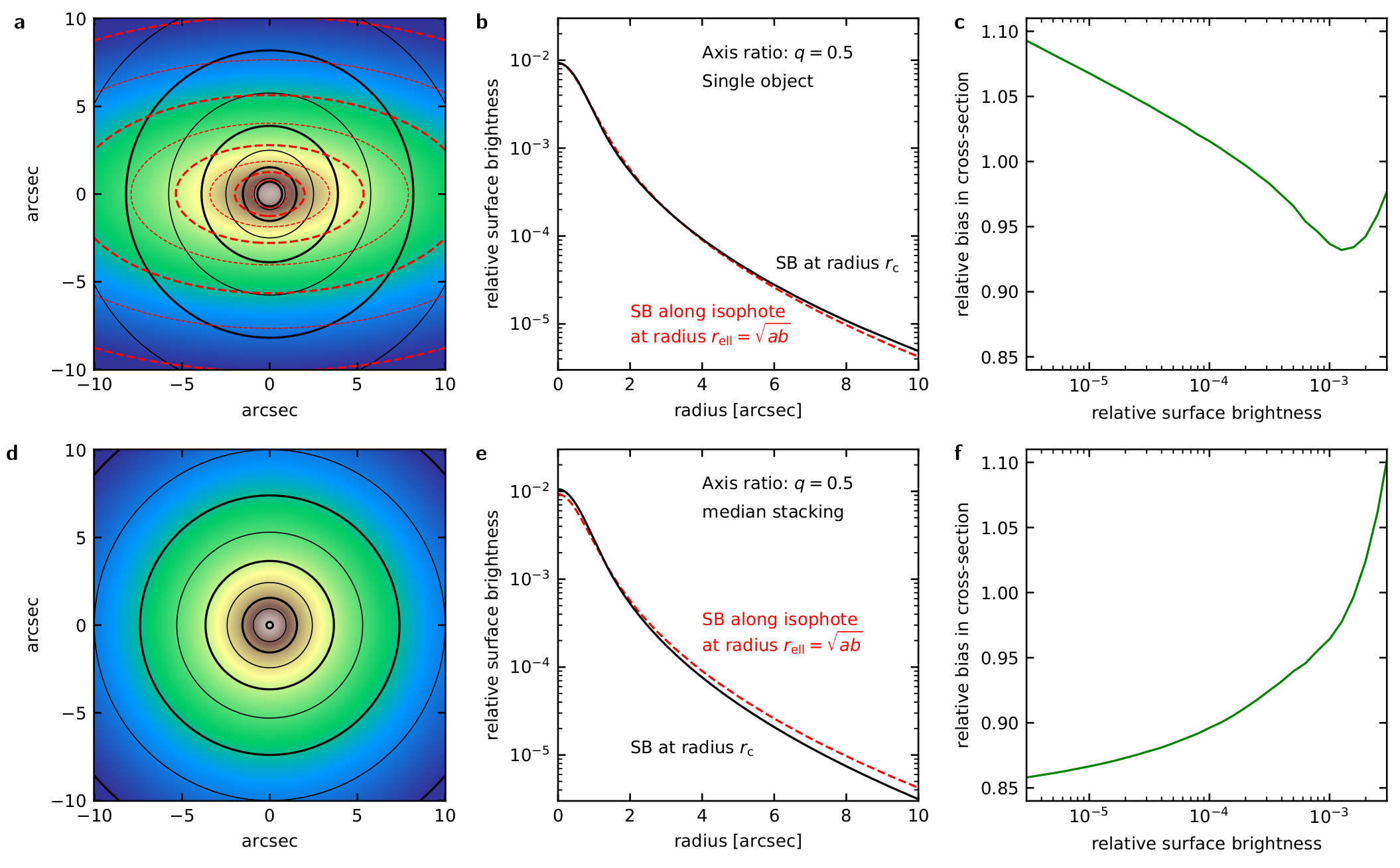}
\caption[]{\textbf{Bias of estimated cross-sections if the emission is non-axisymmetric.} \textbf{a,} Model image of an elongated surface brightness distribution, normalised to an integrated flux of 1, following an elliptical Sersic law with axis ratio $q=0.5$ and smoothed with a Gaussian of 0\farcs8 FWHM. The red-dotted contours trace the isophotes at 0.5~dex separation. The black circles represent the radii where an azimuthally averaged profile over circular annuli gives the same surface brightness values as the corresponding isophote. 
\textbf{b,} Radial profiles of the model image. The red-dotted line represents the input surface brightness law as a function of generalised radius $r_\mathrm{ell}=\sqrt{ab}$ where $a$, $b$ are the major and minor axes of an isophote. The black line shows the profile obtained from azimuthal averaging over circular annuli against radius $r_\mathrm{c}$. 
\textbf{c,} Ratios between the true isophotal cross-sections $\pi a b$ and those estimated from the circularised profile as $\pi r_\mathrm{c}^2$  (i.e., the ratios of the areas of the black circles and the corresponding red-dotted ellipses in panel b), as a function of surface brightness. \textbf{d,} Median-stacked image of an ensemble of 180 model objects with properties each as in panel a, but rotated in position angle between 0\degr\ and 180\degr\ in steps of 1\degr. The black circles show the resulting isophotes at 0.5~dex separation. \textbf{e,} The black line traces the radial profile of the median-stacked image in panel d. The red-dotted line is the true elliptical surface brightness distribution of a single object (same as in panel b). \textbf{f,} Ratios of 
cross-sections obtained from the median stack to the true isophotal ones in a single image, as a function of surface brightness. }
\label{fig:ell}
\end{figure}

\clearpage

\begin{table}
\caption[]{\textbf{Values of the best-fit parameters for the analytic profiles,} obtained from applying GALFIT to the median-stacked images. The first three parameters characterise the circular Sersic model used to describe the Ly$\alpha$ haloes, for each of the three adopted redshift ranges: halo flux $F_\mathrm{h}$ (in $10^{-20}$ erg s$^{-1}$ cm$^{-2}$), effective radius $r_{\mathrm{eff, h}}$ (in arcsec), and Sersic index $n_{\mathrm{h}}$ (dimensionless), followed by the flux of the point-like component $F_\mathrm{ps}$ (same units as $F_\mathrm{h}$). The quoted errors are 1$\sigma$ uncertainty estimates. The last column provides the seeing (FWHM of the mean PSF in arcsec) at the appropriate wavelengths.}
\begin{center}
\begin{tabular}{c r@{\hspace{0.2em}$\pm$\hspace{0.2em}}l r@{\hspace{0.2em}$\pm$\hspace{0.2em}}l r@{\hspace{0.2em}$\pm$\hspace{0.2em}}l r@{\hspace{0.2em}$\pm$\hspace{0.2em}}l c}
\hline\noalign{\smallskip} 
\multicolumn{1}{c}{$z$ range} & \multicolumn{2}{c}{$F_{\mathrm{h}}$} & \multicolumn{2}{c}{$r_{\mathrm{eff, h}}$} & \multicolumn{2}{c}{$n_{\mathrm{h}}$} & \multicolumn{2}{c}{$F_{\mathrm{ps}}$} & \multicolumn{1}{r}{FWHM$_{\mathrm{PSF}}$} \\
\noalign{\smallskip} \hline\noalign{\smallskip} 
3--4 & 1488 & 83 &  0.86 & 0.11 & 2.8 & 1.1 & 232  & 50 & 0.703 \\
4--5 & 931  & 82  &  0.90 & 0.18 & 3.3 & 1.9 & 150  & 40 & 0.654 \\
5--6 & 1002 & 164 & 1.67 & 0.86 & 6.5 & 5.1 & 167 & 34 & 0.606  \\
\noalign{\smallskip} \hline
\end{tabular}
\end{center}
\label{tab:fitpar}
\end{table}

\end{document}